\documentclass[aps,twocolumn,prb,10pt, floatfix,superscriptaddress,longbibliography,final]{revtex4-1}
\usepackage[caption=false]{subfig}
\usepackage{graphicx, amsmath, bm, amsfonts, amssymb, color}
\usepackage[%
  pdftex,
  breaklinks=true,
  colorlinks=true,
  linkcolor=blue,
  urlcolor=blue,
  citecolor=blue
]{hyperref}
\usepackage{verbatim}
\usepackage{xcolor}
\usepackage[braket, qm]{qcircuit}

\newcommand{\RdS}[1]{\color{black}#1\color{black}}

\newcommand{\beq}[0]{\begin{equation}}
\newcommand{\eeq}[0]{\end{equation}}

\begin{document}

\title{Confined Magnons}
\author{Seamus Beairsto}
\affiliation{Department of Physics and Astronomy, University of Victoria, Victoria, British Columbia V8W 2Y2, Canada}
\affiliation{Centre for Advanced Materials and Related Technology, University of Victoria, Victoria, British Columbia V8W 2Y2, Canada}
\author{Maximilien Cazayous}
\affiliation{Laboratoire Mat\'{e}riaux et Ph\'{e}nom\`{e}nes Quantiques, UMR 7162 CNRS, Universit\'{e} Paris Diderot, B\^{a}timent Condorcet 75205 Paris Cedex 13, France}
\author{Randy S. Fishman}
\affiliation{Materials Science and Technology Division, Oak Ridge National Laboratory, Oak Ridge, Tennessee 37830, USA}
\author{Rog\'{e}rio \surname{de Sousa}}
\affiliation{Department of Physics and Astronomy, University of Victoria, Victoria, British Columbia V8W 2Y2, Canada}
\affiliation{Centre for Advanced Materials and Related Technology, University of Victoria, Victoria, British Columbia V8W 2Y2, Canada}

\begin{abstract}
Magnetic structures are known to possess magnon excitations confined to their surfaces and interfaces, but these spatially localized modes are often not resolved in spectroscopy experiments.  We develop a theory to calculate the confined magnon spectra and its associated spin scattering function, which is the physical observable in neutron and electron scattering, and a proxy for \RdS{photon spectroscopy based on X-ray, Raman and THz sources}. 
We show that extra anisotropy at the surface or interface plays a key role in magnon confinement. We obtain analytical expressions for the confinement length scale, and show that it is qualitatively similar for ferromagnets and antiferromagnets in dimension $d\geq 2$. For $d=1$ we find remarkable differences between ferromagnetic and antiferromagnetic models. The theory indicates the presence of several confined magnon resonances in addition to the usual magnons thought to explain the excitations of magnetic nanostructures. Detecting these modes may elucidate the impact of the interface on spin anisotropy and magnetic order. 
\end{abstract}

\maketitle

\section{Introduction}

The excitations of the magnetic state of large (bulk) magnets are well understood to be collective delocalized spin waves, whose quanta are called magnons. These modes are described by continuous dispersion relations that depend on the nature of the magnetic state, i.e. whether the material is a ferromagnet or antiferromagnet.\cite{Demokritov2012, Rezende2019}

In small systems such as e.g. molecular magnets with $N\sim 10$ spins, the magnon approach breaks down.\cite{Dreiser2010, Chiesa2017} 
This raises the question whether magnons in larger ``mesoscopic'' systems 
such as magnetic nanoparticles and few-monolayer thin films can be fully described with the magnon picture.
Comparison between theory and experiment in magnetic structures such as nanoparticles\cite{Hansen2000, Etz2015, Lefmann2015} is quite difficult due to a variety of finite size effects including large surface to volume ratio, lower symmetry, and size/shape distribution.\cite{Feygenson2011, Aupiais2020}  As a result, there is a notable gap in our understanding of magnetic excitations in nanostructures and related confined magnetic systems. 

The breakdown of space translational symmetry in finite magnets produces localized excitations concentrated at the borders of the system, the so called magnon confinement phenomena.\cite{Mathieu1998, Hillebrands2000, Demokritov2001, Park2002, Wieser2008} Confined magnons were traditionally described using macroscopic theory involving simultaneous solution of the Landau-Lifshitz and Maxwell's equations in the magnetostatic limit.\cite{Walker1957, Damon1961, Hillebrands2000, Demokritov2001} This approach involves approximations that are known to become invalid in the large energy and wavevector regime, when the magnon dispersion is dominated by magnetocrystalline anisotropy and exchange interactions.\cite{Erickson1991} 
Moreover, the macroscopic treatment becomes questionable in the presence of discontinuities in spin Hamiltonian parameters, e.g. when there is extra magnetocrystalline anisotropy at the surface or interface. 

This case is of importance to e.g. the research area of magnonics, where one uses few monolayer nanostructures to engineer the desired magnon spectra.\cite{Demokritov2012,Krawczyk2014}
While confined magnons are frequently observed 
in large ferromagnets with Brillouin spectroscopy,\cite{Hillebrands2000, Demokritov2001} and in ultra-thin films with spin-polarized electron energy loss spectroscopy (SPEELS),\cite{Prokop2009,Chen2017,Qin2019,Zakeri2021}
they are frequently ignored in a wide variety of systems, most notably antiferromagnets. To the best of our knowledge, there are no measurements of confined magnons in antiferromagnets.

Here we formulate a theory of confined magnons and their spectroscopy. We present exact analytical results for infinite systems with borders to establish the conditions for existence of confined magnons in simple ferromagnetic and antiferromagnetic models, and then move on to present a general theory that allows the determination of the confined and deconfined spectrum for all finite models. Our method allows explicit numerical computations of the spin scattering function (also called Bloch spectral function), which is directly measurable in spin-polarized neutron\cite{Fishman2018} and electron\cite{Zhang2012} scattering experiments, and is a proxy for 
\RdS{photon spectroscopy based on THz,\cite{Fishman2018} Raman,\cite{Aupiais2020, Cottam1986} and X-rays.\cite{Suzuki2019,Pelliciari2021}}


A key result of our theory is the realization that extra magnetocrystalline anisotropy at the surface or interface plays a crucial role in magnon confinement. 

As pointed out by N\'{e}el,\cite{Neel1954} magnetocrystalline anisotropy for an ion at the surface is different from an ion in the bulk because the former has lower point group symmetry.  This effect is specially important in materials with low magnetocrystalline anisotropy in bulk, such as ferromagnetic iron (Fe). While the bulk anisotropy in Fe is quite low ($2.1\times 10^{-3}$~meV per spin),\cite{Daalderop1990} measurements of the magnon dispersion for a single Fe monolayer on W(110) have determined interface anisotropy per spin to be equal to $K_s=2.3\pm 1.3$~meV,\cite{Prokop2009} $10^3$ times larger than the bulk value. However, we are not aware of measurements of surface/interface anisotropy for antiferromagnets. Recently, the presence of extra magnetic anisotropy in the surface of multiferroic nanoparticles was shown to have huge impact on its antiferromagnetic-ferroelectric state. The nanoparticle magnetic and ferroelectric moments become bistable, enabling the design of ideal memory bits that can be switched electrically and read out magnetically.\cite{Allen2019} While surface anisotropy has been measured in ferromagnets with scanning tunneling microscopy\cite{Rau2014} and Brillouin spectroscopy,\cite{Hillebrands2000} its impact on the magnetic excitations of antiferromagnets has not been studied to date.
It is desirable to identify the spectroscopic signature of surface/interface anisotropy so that the impact of surrounding materials on the magnetic properties of nanostructures can be understood.

We organize the article as follows. Section~\ref{section:analytic} presents \RdS{exact}~analytic theory for confined \RdS{and propagating}~magnons at the surface or interface of \RdS{semi-infinite}~systems with simple model Hamiltonians. We establish the conditions for existence of spatially-localized modes and show that surface/interface anisotropy plays a key role in determining their confinement length scale. We also give estimates for a few common materials. Section~\ref{section:general} presents a general theory applicable to arbitrary models that allows numerical determination of the confined spectra and its associated spin-scattering function. We present explicit numerical calculations for one-dimensional ferro and antiferro systems, allowing visualization of the confined and deconfined spectra, and we compare to analytic theory. Section~\ref{section:conclusions} presents our conclusions.

\begin{figure}
	\centering
	\includegraphics[width=0.5\textwidth]{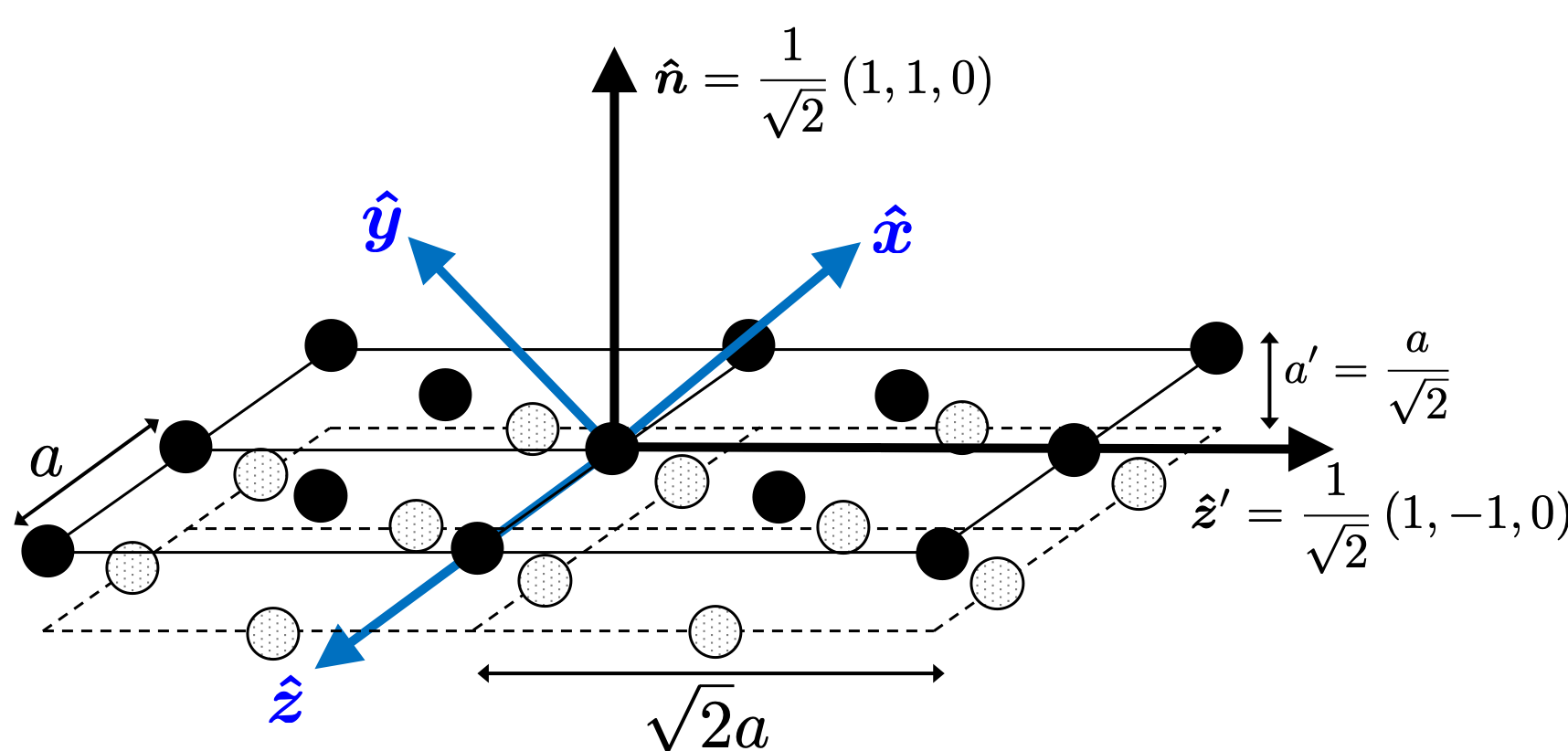}
	\caption{Depicts the top two monolayers of the bcc(110) surface, normal to $\bm{\hat{n}}=(1,1,0)/\sqrt{2}$. Unit vector $\bm{\hat{z}}'$ represents the easy axis direction for magnetocrystalline anisotropy. For Fe(110) this is given by $\bm{\hat{z}}'=(1,-1,0)/\sqrt{2}$. Unit vectors $\bm{\hat{x}}, \bm{\hat{y}}, \bm{\hat{z}}$  represent the usual cubic axes, and $a'$ is the separation between monolayers.} 
	\label{fig:FeSurface}
\end{figure}

\section{Confined magnons at the surface/interface of semi-infinite systems \label{section:analytic}}

We focus on the model Hamiltonian 
\begin{equation}
{\cal H} = \frac{J}{2} \sum_{j} \sum_{\bm{v}_j}\bm{s}_j \cdot \bm{s}_{j+\bm{v}_j} - \sum_{j}K_j(\bm{s}_{j}\cdot \bm{\hat{z}}')^{2},
\label{Hinfty}
\end{equation}
where $J$ is the exchange interaction between nearest-neighbor (n.n.) spins, and $\bm{s}_j$ denotes a spin operator \RdS{in site $j$ of a semi-infinite lattice where half the space is empty or filled with a nonmagnetic material (i.e. the system has a surface in the former case and an interface in the latter). For compact notation
we label each lattice site by $j=(j_1,j_2,j_3)$. This corresponds to location $\bm{r}_j=\sum_i j_i \bm{a}_i$ in the lattice (e.g. $\bm{a}_1=a\bm{\hat{x}}, \bm{a}_2=a\bm{\hat{y}}, \bm{a}_3=a\bm{\hat{z}}$ for the simple cubic lattice)}. 
For each $j$, the set of vectors $\{\bm{v}_j\}$ link the $j$-th spin to its n.n. spins; this set depends on $j$ because spins located at the surface or interface have a reduced number of n.ns.

Model parameter $K_j$ describes the magnetic anisotropy with easy axis along unit vector $\bm{\hat{z}}'$.
\RdS{In this work we take each $K_j$ to assume one of two possible values.  We asssume $K_j= K_s$ for $j\in {\cal I}_n$, where ${\cal I}_n$ denotes an interface or surface normal to unit vector $\hat{\bm{n}}$. For the other spins $j\not\in {\cal I}_n$ we assume $K_j = K$}.
Note that $\hat{\bm{n}}$ may not coincide with the easy axis direction $\bm{\hat{z}}'$. However, we assume the easy axis for spins at the surface/interface to be the same as the easy axis for spins in the bulk. Figure~\ref{fig:FeSurface} illustrates the case of Fe(110). 

In this way $\Delta K_s\equiv (K_s-K)$ models the impact of a nonmagnetic interface. We emphasize that $|\Delta K_s/J|$ can be quite large even for a surface, which corresponds to an interface between vacuum and the magnetic system.\cite{Rau2014} 

We remark that Hamiltonian~(\ref{Hinfty}) does not include long range dipolar interactions. This is not a problem for antiferromagnets since dipolar interactions can be suitably included as an additional contribution to bulk anisotropy $K$, of the order of $K_d= \mu_0 \left(g\mu_B\right)^{2}/v_0\sim 0.1$~meV, where $v_0$ is the unit cell volume and $\mu_B$ is the Bohr magneton.\cite{Hutchings1970}

For ferromagnets, dipolar interactions become important at the low energy and small wavevector regime. Therefore, keep in mind that Hamiltonian~(\ref{Hinfty}) provides a proper description of FMs only when either $K\gtrsim K_d$ (lowest magnon has energy larger than dipolar), or $qa> \sqrt{K_d/(z|J|)}\sim 0.01\pi$ (magnon dispersion dominated by exchange).\cite{Erickson1991} 
\RdS{We remark that previous macroscopic theories\cite{Walker1957, Damon1961, Hillebrands2000, Demokritov2001} are not valid in this regime.
The regime of our theory allows the description of a large number of spectroscopy experiments, including electron,\cite{Prokop2009} neutron,\cite{Loong1984} \RdS{and X-ray\cite{Pelliciari2021}}~scattering, as well as optical spectroscopy based on Raman technique\cite{Aupiais2020} and THz sources.\cite{Fishman2018}} 

In this section, we obtain magnon modes using the method of the classical equations of motion.\cite{Landau1980} This is done by defining a mean-field Hamiltonian ${\cal H}_{MF}$ which replaces $\bm{s}_j$ in Eq.~(\ref{Hinfty}) by its average $\langle\bm{s}_j\rangle$, leading to the system of $N$ coupled equations of motion, one equation for each $j$:
\begin{equation}
    \frac{\partial \langle\bm{s}_j\rangle}{\partial t}= \frac{1}{\hbar}\frac{\partial {\cal H}_{MF}}{\partial \langle\bm{s}_j\rangle}\times \langle\bm{s}_j\rangle,
\label{classicalEOM}
\end{equation}
where each spin precesses about its own local field given by
\begin{equation}
    \frac{\partial {\cal H}_{MF}}{\partial \langle\bm{s}_j\rangle} = J\sum_{\bm{v}_j}\langle\bm{s}_{j+\bm{v}_j}\rangle-2K_j \langle s_{j}^{z}\rangle\hat{\bm{z}}'.
\label{local_field}
\end{equation}
The magnon frequencies in the classical method are identical to the ones obtained by the quantum Holstein-Primmakoff method that we describe below for evaluation of the spin scattering function, within the approximation of negligible magnon-magnon interactions. However, as we show here the classical method leads to \RdS{exact analytical solutions for confined and propagating magnon modes in the presence of an interface}.

\begin{figure}
	\centering
	\includegraphics[width=0.5\textwidth]{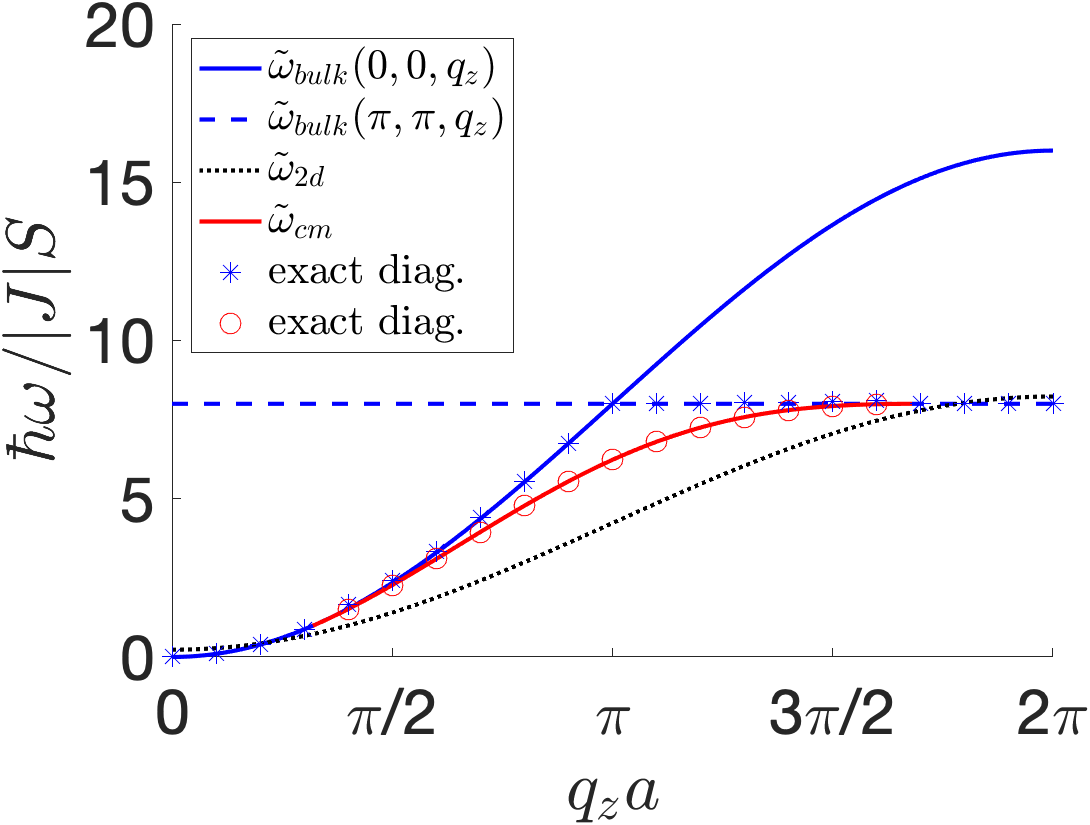}
	\caption{The red (solid) and blue (solid, dashed) curves depict the lowest magnon modes \RdS{using the analytical dispersions Eqs.~(\ref{cmFM})~and~(\ref{bulkFM})}~for the (110) interface of the bcc lattice, as a function of $\bm{q}_{\parallel}=q_z\bm{\hat{z}}$.  The choice of parameters describe the Fe/W(110) interface: $K_s=2.3$~meV,\cite{Prokop2009} $J=20$~meV,\cite{Loong1984} and 
	$K=2.1\times 10^{-3}$~meV.\cite{Daalderop1990} The stars and circles were obtained by exact diagonalization of Eqs.~(\ref{mixedEOM_FM}) for 21 monolayers. These are seen to agree perfectly with the analytical dispersions for the confined magnon (red-solid, Eq.~(\ref{cmFM})) and bulk magnon (blue-solid and dashed, Eq.~(\ref{bulkFM})). \RdS{This shows that the exponential trial Eq.~(\ref{ansatzFM}) for the confined magnon is exact}. Note how the confined magnon only exists in the range $0.30 \pi < q_z a < 1.7\pi$; outside this range its frequency merges with the bulk modes and the inverse length scale $\kappa_\perp$ becomes negative, signaling the nonexistence of confined modes. For comparison the black-dotted curve shows the dispersion of a single Fe(110) monolayer with the same parameters.} 
	\label{Fe110}
\end{figure}

\subsection{Ferromagnetic models \label{subsection:AnalyticFM}}

For $J=-|J|<0$ and $K\geq 0$ the ground state of Eq.~(\ref{Hinfty}) is a homogeneous ferromagnet (FM) with $\langle\bm{s}_j\rangle=S\bm{\hat{z}}'$ for all $j$, where $S$ is the spin quantum number. This is the case provided that the interface anisotropy is not too negative, $\Delta K_s\equiv K_s-K\geq \Delta K_c$ where $\Delta K_c<0$ is a critical value to be determined below. When $\Delta K_s<\Delta K_c$ a spin-flip transition occurs leading to interface spins pointing in a different direction than bulk spins.\cite{Levy1981}

The magnon modes are obtained by plugging $\langle \bm{s}_j(t)\rangle = S\bm{\hat{z}}'+\delta \bm{s}_j e^{-i\omega t}$ into Eq.~(\ref{classicalEOM}) with $\delta\bm{s}_j\perp \bm{\hat{z}}'$. After linearization and changing variables to $\delta s^{+}_{j}=(\bm{\hat{x}}'+i \bm{\hat{y}}')\cdot\delta\bm{s}_{j}$, where $\bm{\hat{x}}',\bm{\hat{y}}'$ forms a set of axes perpendicular to $\bm{\hat{z}}'$ we get 
\begin{equation}
    \left[\tilde{\omega}-z_j-\frac{2K_j}{|J|}\right]\delta s^{+}_{j}+\sum_{\bm{v}_j}\delta s^{+}_{j+\bm{v}_j}=0,
\label{classicalLEOM}
\end{equation}
where $\tilde{w}=\hbar\omega/(|J|S)$, and $z_j=\sum_{\bm{v}_j}(1)$ is the number of n.ns. for spin $j$. 

For simplicity we specialize to the case where ${\cal I}_n$ corresponds to a plane of inversion symmetry of the lattice, so that for $j\in {\cal I}_n$ the the set of missing $\bm{v}_j$'s is equal to $\{-\bm{v}_{j}\}$. This case includes all high symmetry planes of the sc (simple cubic), bcc (body-centered cubic), and fcc (face-centered cubic) lattices. In all these lattices we can choose primitive vectors $\bm{a}_1,\bm{a}_2$ in the plane so that it is convenient to define mixed coordinates by taking the Fourier transform over the spins \RdS{in the planes parallel to}~${\cal I}_n$,
\begin{equation}
    \delta s^{+}_{\bm{q}_\parallel,j_3} = \frac{1}{\sqrt{N_{\parallel}}}\sum_{j_1,j_2} e^{i\bm{q}_{\parallel}\cdot \bm{r}_{j}} \delta s^{+}_{j},
\end{equation}
where $\bm{q}_{\parallel}$ is a real vector perpendicular to $\hat{\bm{n}}$, and $j_3=0,1,\ldots$ labels the number of monolayers \RdS{away}~from the interface. 
Equation~(\ref{classicalLEOM}) becomes 
\begin{eqnarray}
&&\left\{\tilde{\omega}-z_{\perp}-z_{\parallel}\left[1-\gamma_{\parallel}(\bm{q}_{\parallel})\right]-\frac{2K}{|J|}\right.\nonumber\\
&&\left.-\left(\frac{2\Delta K_s}{|J|}-\frac{z_{\perp}}{2}\right)\delta_{j_3,0}\right\}
\delta s^{+}_{\bm{q}_{\parallel},j_3}\nonumber\\
&&+\frac{z_{\perp}}{2}\gamma_{\perp}(\bm{q}_{\parallel})\left[\left(1-\delta_{j_3,0}\right)\delta s^{+}_{\bm{q}_{\parallel},j_3-1}+\delta s^{+}_{\bm{q}_{\parallel},j_3+1}\right]=0.
\label{mixedEOM_FM}
\end{eqnarray}
These equations are obtained by separating n.n. vectors into two disjoint sets, $\{\bm{v}_j\}=\{\bm{v}_{j\parallel}\}\cup \{\bm{v}_{j\perp}\}$, where 
$\bm{v}_{j\parallel}\cdot \hat{\bm{n}}=0$ and $\bm{v}_{j\perp}\cdot \hat{\bm{n}}\neq 0$. As a result we can write 
$z_{j_3}=z_{\parallel}+(2-\delta_{j_3,0})z_{\perp}/2$, where
$z_{\parallel}=\sum_{\bm{v}_{j_3\parallel}}(1)$, and $z_{\perp}=\sum_{\bm{v}_{j_3\perp}}(1)$ \RdS{are defined for $j_3> 0$}~(outside ${\cal I}_n$). 
This separation naturally leads to two kinds of dispersion functions,
\begin{equation}
\gamma_{\parallel}(\bm{q}_{\parallel})=\frac{1}{z_{\parallel}}\sum_{\bm{v}_{j\parallel}}e^{i\bm{q}_{\parallel}\cdot \bm{v}_{j\parallel}},\;\;
\gamma_{\perp}(\bm{q}_{\parallel})=\frac{1}{z_{\perp}}\sum_{\bm{v}_{j\perp}}e^{i\bm{q}_{\parallel}\cdot \bm{v}_{j\perp}},
\label{gamma_par_perp}
\end{equation}
again defined for $j\not\in {\cal I}_n$ (See Table~\ref{TableLatticeParameters}). 

We now search for confined magnons in Eqs.~(\ref{mixedEOM_FM}) by plugging the \RdS{pure confined magnon}~trial solution
\begin{equation}
    \delta s^{+}_{\bm{q}_{\parallel},j_3}= e^{\left(iQ_{\perp}-\kappa_{\perp}\right)a'j_3}\delta s^{+}_{\bm{q}_{\parallel}},
    \label{ansatzFM}
\end{equation}
where $Q_{\perp}$ and $\kappa_{\perp}$ are real, $a'$ is the separation between two monolayers along $\hat{\bm{n}}$, and $\delta s^{+}_{\bm{q}_{\parallel}}$ is an arbitrary amplitude.
Note that in Eq.~(\ref{ansatzFM}) $\kappa_\perp$ plays the role of an inverse length scale for confinement. We must have $\kappa_{\perp}\geq 0$, otherwise the modulus of Eq.~(\ref{ansatzFM}) blows up at large $j_3$.

The trial solution reduces the system of Eqs.~(\ref{mixedEOM_FM})  to only two equations, one for $j_3=0$ ($j\in {\cal I}_n$) and another for $j_3\neq 0$:
\begin{subequations}
\begin{eqnarray}
 \tilde{\omega}&=&\frac{2K_s}{|J|}+z_{\parallel}\left[1-\gamma_{\parallel}(\bm{q}_{\parallel})\right]+\frac{z_{\perp}}{2}\left[1-\gamma_{\perp}(\bm{q}_{\parallel})\right.\nonumber\\
 &&\times \left. e^{iQ_{\perp}a'}e^{-\kappa_{\perp} a'}\right],\label{jI}\\
 \tilde{\omega}&=&\frac{2K}{|J|}+z_{\parallel}\left[1-\gamma_{\parallel}(\bm{q}_{\parallel})\right]+z_{\perp}\left[1-\frac{\gamma_{\perp}(\bm{q}_{\parallel})}{2}\right.\nonumber\\
 &&\times \left.\left(e^{iQ_{\perp}a'}e^{-\kappa_{\perp} a'}+e^{-iQ_{\perp}a'}e^{\kappa_{\perp} a'}\right)\right].\label{jnotI}
\end{eqnarray}
\label{jIcombined}
\end{subequations}

Subtracting Eq.~(\ref{jnotI}) from Eq.~(\ref{jI}) and simplifying we get
\begin{equation}
    \kappa_{\perp}a' = \ln{\left[\frac{e^{iQ_{\perp}a'}}{\gamma_{\perp}(\bm{q}_{\parallel})}\left(1-\frac{4(\Delta K_s)}{z_{\perp}|J|}\right)\right]}.
\label{kappa_perp}
\end{equation}
Confined magnons exist when Eq.~(\ref{kappa_perp}) admits solutions with $\kappa_{\perp}>0$ (localized in space) and $\tilde{\omega}\geq 0$ (no spin-flip instability). For $\bm{q}_{\parallel}=\bm{0}$, $\gamma_{\perp}=1$. Hence, the modulus of the argument of the $\ln{}$ is less than $1$ whenever $0\leq \Delta K_s/|J|\leq z_{\perp}/2$. In this case $\kappa_{\perp}$ is necessarily negative, so no confined magnon with $\bm{q}_{\parallel}=\bm{0}$ exists for $\Delta K_s$ in this range. 

However, for nonzero wavevectors in the range $0<q_{c1}<q_{\parallel}<q_{c2}$, $\gamma_{\perp}(\bm{q}_{\parallel})<1$ becomes small enough so that the modulus of the argument of the $\ln{}$ is greater than $1$ for any value of $\Delta K_s$. This shows that confined magnons may exist at nonzero wavevectors, even when $0\leq \Delta K_s/|J|\leq z_{\perp}/2$. 

In contrast, for $\Delta K_s/|J|>z_{\perp}/2$ the choice $Q_{\perp}a'=\pi$ makes $\kappa_{\perp}$ and $\tilde{\omega}$ always positive, so a confined magnon solution exists for all $\bm{q}_{\parallel}$, \emph{including $q_{\parallel}=0$}. Also, when  $\Delta K_s<0$ we can make $\kappa_{\perp}>0$ by choosing $Q_{\perp}=0$, but here $\tilde{\omega}$ can become negative if $\Delta K_s$ is too negative, so the confined magnon will exist in the range $\Delta K_c\leq \Delta K_s<0$ for all $\bm{q}_{\parallel}$. 

When it exists the confined magnon has inverse length scale given by
\begin{equation}
    \kappa_{\perp}=\frac{1}{a'}\ln{\left|\frac{1}{\gamma_{\perp}(\bm{q}_{\parallel})}\left[1-\frac{4(\Delta K_s)}{z_{\perp}|J|}\right]\right|},
\label{kappa_perp_final}
\end{equation}
and plugging this into Eq.~(\ref{jnotI}) we get the confined magnon frequency:
\begin{eqnarray}
    \frac{\hbar\omega^{FM}_{cm}(\bm{q}_{\parallel})}{|J|S}&=&\frac{2K}{|J|}+z_{\parallel}\left[1-\gamma_{\parallel}(\bm{q}_{\parallel})\right]\nonumber\\
    &&+\frac{z_{\perp}}{2}\left\{
    \frac{1-\left(\frac{4\Delta K_s}{z_{\perp}|J|}\right)^{2}-\left[\gamma_{\perp}(\bm{q}_{\parallel})\right]^{2}}{1-\frac{4\Delta K_s}{z_{\perp}|J|}}
    \right\}.
    \label{cmFM}
\end{eqnarray}
\RdS{We emphasize that no approximation was used to obtain this expression; it followed from the trial solution Eq.~(\ref{ansatzFM})}.

\RdS{In addition to confined modes, Eq.~(\ref{mixedEOM_FM}) also contains solutions for propagating modes. These can be obtained by plugging the pure bulk magnon trial solution
\begin{equation}
    \delta s^{+}_{\bm{q}_{\parallel},j_3}=\sin{\left(Q_{\perp}a'j_3+\phi\right)}\delta s^{+}_{\bm{q}_{\parallel}},
\end{equation}
with $\phi\in [0,\pi)$ a phase shift. This trial again reduces the system Eq.~(\ref{mixedEOM_FM}) to only two equations; the one for $j_3\geq 1$ is the usual dispersion for a FM with full translation invariance,\cite{Kittel1987}
\begin{equation}
    \frac{\hbar\omega^{FM}_{bulk}(\bm{q})}{|J|S}= \frac{2K}{|J|}+z\left[1-\gamma(\bm{q})\right],
\label{bulkFM}
\end{equation}
where $z=z_{\parallel}+z_{\perp}$ is the number of n.ns. and $\gamma(\bm{q})=\frac{1}{z}\sum_{\bm{v}}e^{i\bm{q}\cdot\bm{v}}$ is the bulk dispersion function. 
The equation for $j_3=0$ determines the phase shift,
\begin{equation}
    \cot{(\phi)}=-\frac{1-\frac{4(\Delta K_s)}{z_{\perp}|J|}-\gamma_{\perp}(\bm{q}_{\parallel})\cos{(Q_{\perp}a')}}{\gamma_{\perp}(\bm{q}_{\parallel})
    \sin{(Q_{\perp}a')}}.
\label{phase_shift1}
\end{equation}
Once again, these solutions are exact. 

Rearrange Eq.~(\ref{phase_shift1}) to get
\begin{equation}
    \frac{\sin{(\phi-Q_{\perp}a')}}{\sin{(\phi)}}=\frac{1}{\gamma_{\perp}(\bm{q}_{\parallel})}\left[1-\frac{4(\Delta K_s)}{z_{\perp}|J|}\right]. 
\end{equation}
When $Q_{\perp}=0$ ($\pi$), the LHS equals $+1$ ($-1$) for all $\phi$; hence, when the confined mode exists (argument of $\ln$ greater than $1$ in Eq.~(\ref{kappa_perp_final2})), a solution for $\phi$ can not be found}.

The confined magnon frequency Eq.~(\ref{cmFM}) must be positive for the homogeneous FM state to be stable;
\RdS{the criteria can be found by setting $q_{\parallel}=0$ and $\omega^{FM}_{cm}=0$ in Eq.~(\ref{cmFM}) and solving for $\Delta K_s$:}
\begin{equation}
    \Delta K_s\geq \Delta K_c = -\frac{K}{2}\left(1+\sqrt{1+\frac{z_{\perp}|J|}{K}}\right).
\label{DeltaKc}
\end{equation}
For $\Delta K_s < \Delta K_c$ interface spins point in a different direction than interior spins. The spin order becomes noncollinear;\cite{Levy1981} confined magnons will be present but their frequency is no longer described by Eq.~(\ref{cmFM}). 
\begin{widetext}

\begin{table}
  \begin{center}
\caption{Parameters used to describe cubic lattices with high-symmetry interfaces ${\cal I}_n$. Note that a spin $j\in {\cal I}_n$ has $z_j=z_{\parallel}+z_{\perp}/2$ n.n., while a spin $j\not\in {\cal I}_n$ has $z_j=z_{\parallel}+z_{\perp}$. Parameter $a'$ is the separation between monolayers.}
    \label{TableLatticeParameters}
\vspace{2ex}
    \begin{tabular}{l c c c c c}
\hline\hline
Lattice($\bm{\hat{n}}$) & $z_{\parallel}$ & $z_{\perp}$ &  $a'$ & $\gamma_{\parallel}(\bm{q}_{\parallel})$ & $\gamma_{\perp}(\bm{q}_{\parallel})$ \\
\hline
sc(001)  & 4 & 2 & $a$   & $\frac{1}{2}\left[\cos{(q_xa)}+\cos{(q_ya)}\right]$ & $1$ \\
bcc(001) & 0 & 8 & $a/2$ & $0$ & $\cos{\left(q_x\frac{a}{2}\right)}\cos{\left(q_y\frac{a}{2}\right)}$ \\
bcc(110) & 4 & 4 & $a/\sqrt{2}$ & $\cos{\left[\left(q_x-q_y\right)\frac{a}{2}\right]}\cos{\left(q_z\frac{a}{2}\right)}$ & $\cos{\left(q_z\frac{a}{2}\right)}$ \\
fcc(001) & 4 & 8 & $a/2$ & $\cos{\left(q_x\frac{a}{2}\right)}\cos{\left(q_y\frac{a}{2}\right)}$ & 
$\frac{1}{2}\left[\cos{\left(q_x\frac{a}{2}\right)}+\cos{\left(q_y\frac{a}{2}\right)}\right]$\\

sq(01) & 2 & 2 & $a$   & $\cos{(q_xa)}$ &  $1$\\
chain(edge)  & 0 & 2 & $a$   & $0$ &  $1$\\
\hline\hline
    \end{tabular}
  \end{center}
\end{table}
\begin{table}
  \begin{center}
    \caption{Predicted values for the lowest ($q=0$) bulk and confined magnon energies for \{100\} interfaces and $q_{\parallel}=0$, for antiferromagnets MnF$_2$ and FeF$_2$\cite{Hutchings1970}, and for BiFeO$_3$ nanoparticles assuming homogeneous AFM order.\cite{Aupiais2020, Buhot2015, Matsuda2012} In the absence of measurements of $K_s$ we chose $K_s=K-\varepsilon J$ with $\epsilon J\ll 1$.}
    \label{TableAFM}
\vspace{2ex}
    \begin{tabular}{l c c c c c c}
\hline\hline
 &  Latt. & $S$ & $J$ (meV) & $K$ (meV)  & $\hbar\omega^{AFM}_{bulk}$ (meV) & $\hbar\omega^{AFM}_{cm}$ (meV)  \\
\hline
MnF$_2$   &  bcc & 5/2 & 0.30  & 0.0184  & 1.06 & 0.092 \\
FeF$_2$   &  bcc & 2   & 0.45  & 0.623   & 6.49 & 2.49 \\
BiFeO$_3$ &  sc  & 5/2 & 6.48  & 0.0035   & 1.85 & 1.51 \\
\hline\hline
    \end{tabular}
  \end{center}
\end{table}
\end{widetext}

Figure~\ref{Fe110} presents a comparison between the analytic results of Eqs.~(\ref{cmFM})~and~(\ref{bulkFM}) and exact numerical diagonalization of Eqs.~(\ref{mixedEOM_FM}) for $N_{\perp}=21$ monolayers. 
\RdS{It shows that the trial solution Eq.~(\ref{ansatzFM}) used to obtain Eq.~(\ref{cmFM}) provides the \emph{exact solution} for the confined magnon in a semi-infinite system}.
The calculation is done for the Fe/W(110) interface,
assuming easy axis $\bm{\hat{z}}'=(1,-1,0)/\sqrt{2}$ as shown in Fig.~\ref{fig:FeSurface} and the value $K_s=2.3$~meV  measured with SPEELS,\cite{Prokop2009}
$J=20$~meV from neutron scattering,\cite{Loong1984} and $K=0$ due to the symmetry of Fe's bcc lattice (in bulk Fe the anisotropy is quartic in the spin operators).~For each $\bm{q}_{\parallel}=(0,0,q_z)$, the circles and stars show the two lowest frequency magnons obtained by exact diagonalization. These are seen to agree with the analytic expressions for the confined magnon Eq.~(\ref{cmFM}) and for the usual bulk mode in three dimensions Eq.~(\ref{bulkFM}). Notably, the confined magnon only exists in the range $0.30 \pi < q_z a < 1.7\pi$, when $\kappa_{\perp}>0$ (red curve). 
The confined mode ceases to exist when its frequency becomes greater than either the lowest bulk mode, set by $\tilde{\omega}_{bulk}(0,0,q_z)$ or  
$\tilde{\omega}_{bulk}(\pi,\pi,q_z)=2K/|J|+z$. 

For comparison we also present the $d=2$ (110) monolayer dispersion,
\begin{equation}
    \frac{\hbar\omega^{FM}_{2d}(\bm{q}_{\parallel})}{|J|S}= \frac{2K_s}{|J|}+z_\parallel\left[1-\gamma_{\parallel}(\bm{q}_{\parallel})\right], 
\label{bulk2dFM}
\end{equation}
assuming the same $K_s, J$ as in the $d=3$ case above. Note how the confined magnon dispersion lies in-between the bulk and monolayer dispersions. Figure~\ref{Fe110} is in qualitative agreement with Fig.~3 of Ref.~\onlinecite{Prokop2009} which claimed measurements for the Fe(110)/vaccum confined magnon (24 monolayer sample) together with a single monolayer of Fe/W(110). 

\begin{figure}
	\centering
	\includegraphics[width=0.5\textwidth]{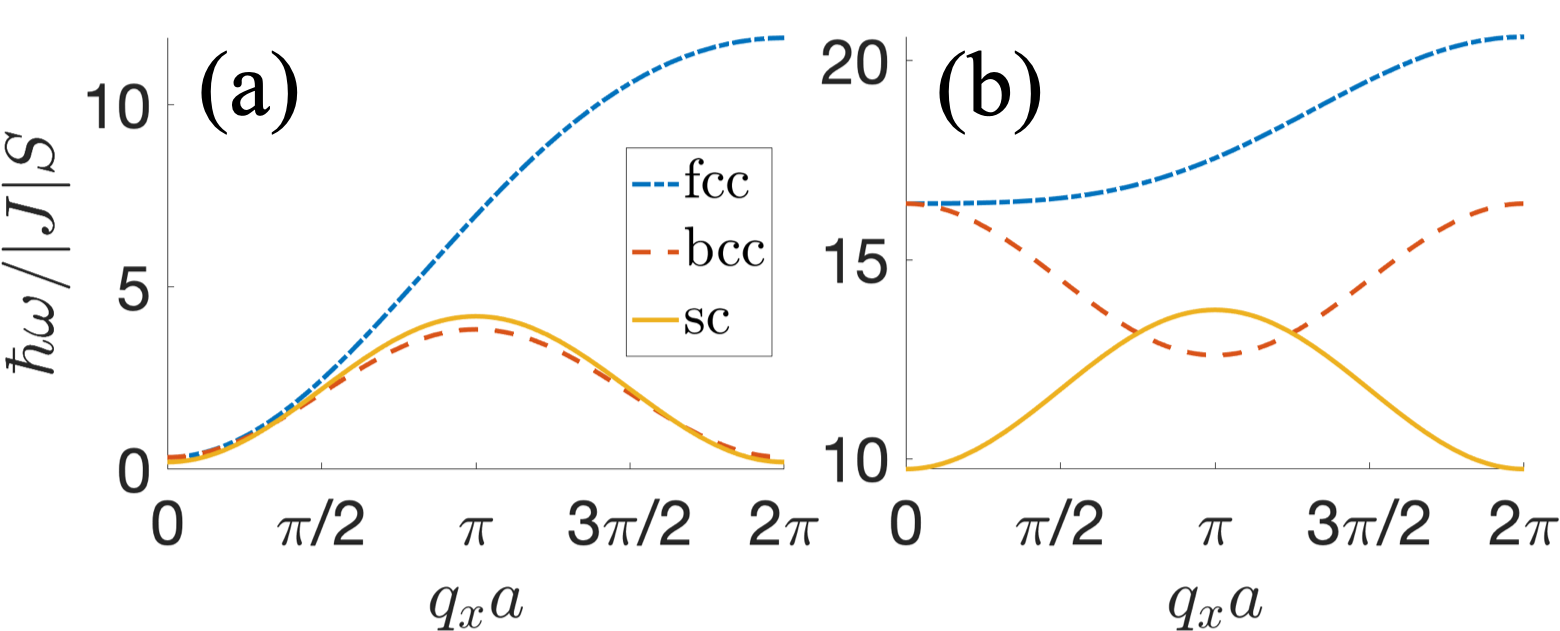}
	\caption{Dispersion for ferromagnetic confined magnons [Eq.~(\ref{cmFM})] for the	family of \{100\} interfaces of three different cubic lattices (face-centered, body-centered, simple cubic), and $\bm{q_{\parallel}}=(q_x,0)$. (a) $K/|J|=0.2$ and $\Delta K_s/|J|=-0.28$; (b) $K/|J|=0.2$ and $\Delta K_s/|J|=4.1$.}
	\label{FM_cm_dispersion}
\end{figure}

Figure~\ref{FM_cm_dispersion} shows the confined magnon dispersion Eq.~(\ref{cmFM}) for the family of \{100\} interfaces in the three cubic lattices calculated with parameters from Table~\ref{TableLatticeParameters}. Figure~\ref{FM_cm_dispersion}(a) with $K/|J|=0.2$, $\Delta K_s/|J|=-0.28$, and Fig.~\ref{FM_cm_dispersion}(b) with $K/|J|=0.2$, $\Delta K_s/|J| =4.1$. For $\Delta K_s <0$ the confined magnon is a low-frequency ``acoustic mode''; in contrast, when  $\Delta K_s/|J| >z_{\perp}/2$ it becomes instead a high-frequency ``optical mode''. For the bcc lattice the lowest optical mode occurs away from the $q_{\parallel}=0$ zone center. 

For the ferromagnetic models, the confined magnon behaves similarly in all dimensions $d=1,2,3$, apart from obvious differences in the dispersion functions Eq.~(\ref{gamma_par_perp}). As we shall see, the situation is quite different for antiferromagnetic models.

\subsection{Antiferromagnetic models in two and three dimensions\label{AFM2d3d}}

For $J>0$ Hamiltonian (\ref{Hinfty}) leads to the homogeneous ``G-type'' antiferromagnetic state 
\begin{equation}
    \langle\bm{s}_j\rangle=(-1)^{\sum_i j_i}S\bm{\hat{z}}',
    \label{gtype}
\end{equation}
provided that $K\geq 0$ and $\Delta K_s$ is not too negative. Assume $\langle \bm{s}_j(t)\rangle = \langle\bm{s}_j\rangle + \delta \bm{s}_j e^{-i\omega t}$ 
with $\delta \bm{s}_j\perp \bm{\hat{z}}'$
and plug into Eq.~(\ref{classicalEOM}) to obtain
\begin{equation}
    \left[\tilde{\omega}(-1)^{\sum_ij_i}-z_j-\frac{2K_j}{J}\right]\delta s^{+}_{j}-\sum_{\bm{v}_j}\delta s^{+}_{j+\bm{v}_j}=0.
\label{AFMclassicalLEOM}
\end{equation}

To proceed we again define mixed $\delta s^{+}_{\bm{q}_{\parallel},j_3}$ coordinates by Fourier transformation on $j_1,j_2$. 
Equation~(\ref{AFMclassicalLEOM}) becomes 
\begin{eqnarray}
    &&\tilde{\omega}(-1)^{j_3}\delta s^{+}_{\bm{q}_{\parallel}+\bm{Q}_{\parallel},j_3}-\left(z_j+\frac{2K_j}{J}\right)\delta s^{+}_{\bm{q}_{\parallel},j_3}\nonumber\\
    &&-\sum_{\bm{v}_{j_3}}e^{-i\bm{q}_{\parallel}\cdot\bm{v}_{j_3}}\delta s^{+}_{\bm{q}_{\parallel},j_3}=0,
\label{mixedEOM}
\end{eqnarray}
where $e^{i\bm{Q}_{\parallel}\cdot \bm{r}_j}=(-1)^{j_1+j_2}$, e.g. 
$\bm{Q}_{\parallel}=(\pi,\pi,0)$ for sc and bcc lattices, $\bm{Q}_{\parallel}=(2\pi,0,0)$ for fcc.

We propose the confined magnon trial solution
\begin{subequations}
\begin{eqnarray}
    \delta s^{+}_{\bm{q}_{\parallel},j_3}&=&e^{-\kappa_\perp a'j_3} \delta s^{+}_{\bm{q}_{\parallel},A}\;{\rm for}\;j_3\;{\rm even},\\
    \delta s^{+}_{\bm{q}_{\parallel},j_3}&=&e^{-\kappa_\perp a'j_3} \delta s^{+}_{\bm{q}_{\parallel},B}\;{\rm for}\;j_3\;{\rm odd},
\end{eqnarray}
\end{subequations}
with different amplitudes $\delta s^{+}_{\bm{q}_{\parallel},A}, \delta s^{+}_{\bm{q}_{\parallel},B}$ for monolayers with $j_3$ even and odd, respectively. Plug this into Eqs.~(\ref{mixedEOM}) and this time they get reduced to three equations for $j_3=0,1,2$:
\begin{widetext}
\begin{subequations}
\begin{eqnarray}
\tilde{w}\delta s^{+}_{\bm{q}_{\parallel}+\bm{Q}_{\parallel},A}&=&\left(z-\frac{z_{\perp}}{2}+\frac{2K_s}{J}+z_{\parallel}\gamma_{\parallel}(\bm{q}_{\parallel})\right)\delta s^{+}_{\bm{q}_{\parallel},A}+\frac{z_{\perp}}{2}\gamma_{\perp}(\bm{q}_{\parallel})e^{-\kappa_{\perp}a'}
\delta s^{+}_{\bm{q}_{\parallel},B},
\label{AFM1}\\
-\tilde{w}\delta s^{+}_{\bm{q}_{\parallel}+\bm{Q}_{\parallel},B}&=&\left(z+\frac{2K}{J}+z_{\parallel}\gamma_{\parallel}(\bm{q}_{\parallel})\right)\delta s^{+}_{\bm{q}_{\parallel},B}+z_{\perp}\gamma_{\perp}(\bm{q}_{\parallel})\cosh{(\kappa_{\perp}a')}
\delta s^{+}_{\bm{q}_{\parallel},A},
\label{AFM2}\\
\tilde{w}\delta s^{+}_{\bm{q}_{\parallel}+\bm{Q}_{\parallel},A}&=&\left(z+\frac{2K}{J}+z_{\parallel}\gamma_{\parallel}(\bm{q}_{\parallel})\right)\delta s^{+}_{\bm{q}_{\parallel},A}+z_{\perp}\gamma_{\perp}(\bm{q}_{\parallel})\cosh{(\kappa_{\perp}a')}
\delta s^{+}_{\bm{q}_{\parallel},B},
\label{AFM3}
\end{eqnarray}
\end{subequations}
\end{widetext}
where $\gamma_{\parallel,\perp}(\bm{q}_{\parallel})$ are defined in Eq.~(\ref{gamma_par_perp}).

Subtract Eq.~(\ref{AFM1}) from (\ref{AFM3}) to get 
\begin{equation}
    \frac{\delta s^{+}_{\bm{q}_{\parallel},A}}{\delta s^{+}_{\bm{q}_{\parallel},B}}=-\frac{\gamma_{\perp}(\bm{q}_{\parallel})e^{\kappa_{\perp}a'}}{1-\frac{4\Delta K_s}{z_{\perp}J}}=-\frac{\delta s^{+}_{\bm{q}_{\parallel}+\bm{Q}_{\parallel},A}}{\delta s^{+}_{\bm{q}_{\parallel}+\bm{Q}_{\parallel},B}},
\label{AFM4}
\end{equation}
where in the last identity we used $\gamma_{\perp}(\bm{q}_{\parallel}+\bm{Q}_{\parallel})=-\gamma_{\perp}(\bm{q}_{\parallel})$. Use Eq.~(\ref{AFM4}) to convert the last term in Eq.~(\ref{AFM2}) into $\delta s^{+}_{\bm{q}_{\parallel},B}$, and plug $\bm{q}_{\parallel}\rightarrow \bm{q}_{\parallel}+ \bm{Q_{\parallel}}$ to obtain a pair of equations coupling $\delta s^{+}_{\bm{q}_{\parallel}+\bm{Q}_{\parallel},B}$ to $\delta s^{+}_{\bm{q}_{\parallel},B}$.  The zero determinant condition then leads to 
\begin{eqnarray}
    \tilde{\omega}^2&=&\left\{z+\frac{2K}{J}-\frac{z_{\perp}[\gamma_{\perp}(\bm{q}_{\parallel})]^2 \left(1+e^{2\kappa_{\perp}a'}\right)}{2\left(1-\frac{4\Delta K_s}{z_{\perp}J}\right)}\right\}^2\nonumber\\
    &&-\left[z_{\parallel}\gamma_{\parallel}(\bm{q}_{\parallel})\right]^2.
    \label{AFM5}
\end{eqnarray}
Similarly, convert Eq.~(\ref{AFM3}) into two equations coupling the A sublattice amplitudes and get
\begin{eqnarray}
    \tilde{\omega}^2&=&\left\{z+\frac{2K}{J}-\frac{z_{\perp}}{2}\left(1-\frac{4\Delta K_s}{z_{\perp}J}\right)
    \left(1+e^{-2\kappa_{\perp}a'}\right)\right\}^2\nonumber\\
    &&-\left[z_{\parallel}\gamma_{\parallel}(\bm{q}_{\parallel})\right]^2.
    \label{AFM6}
\end{eqnarray}
Equations~(\ref{AFM5})~and~(\ref{AFM6}) are equal to each other when $\kappa_{\perp}$ is given by 
\begin{equation}
    \kappa_{\perp}=\frac{1}{a'}\ln{\left|\frac{1}{\gamma_{\perp}(\bm{q}_{\parallel})}\left[1-\frac{4(\Delta K_s)}{z_{\perp}|J|}\right]\right|},
\label{kappa_perp_final2}
\end{equation}
the same found for FMs (see Eq.~(\ref{kappa_perp_final})). Plug this back into Eq.~(\ref{AFM6}) and we get the AFM confined magnon frequency,
\begin{widetext}
\begin{equation}
    \frac{\hbar\omega^{AFM}_{cm}(\bm{q}_{\parallel})}{JS}=\sqrt{\left\{z+\frac{2K}{J}-\frac{z_{\perp}}{2}\left(1-\frac{4\Delta K_s}{z_{\perp}J}\right)
    -\frac{z_{\perp}[\gamma_{\perp}(\bm{q}_{\parallel})]^2}{2\left(1-\frac{4\Delta K_s}{z_{\perp}J}\right)}
    \right\}^2-\left[z_{\parallel}\gamma_{\parallel}(\bm{q}_{\parallel})\right]^2}.
    \label{cmAFM}
\end{equation}
\end{widetext}
The argument of the square root is positive provided that
$\Delta K_s\geq \Delta K_c$ with $\Delta K_c$ given by Eq.~(\ref{DeltaKc}). This shows that the critical value for interface spin-flip transition in AFMs is the same as in FMs. 

Equation~(\ref{cmAFM}) should be compared to the well-known bulk result:
\begin{equation}
    \frac{\hbar\omega^{AFM}_{bulk}(\bm{q})}{JS}= 
    z\sqrt{\left(1+\frac{2K}{zJ}\right)^2-\left[\gamma(\bm{q})\right]^2}.
\label{bulkAFM}
\end{equation}
When $K_s=K-\varepsilon J$ where $\varepsilon$ is small, a confined magnon exists for $q_{\parallel}=0$. The ratio of its frequency to the lowest bulk magnon in the limit $\varepsilon\rightarrow 0+$ is given by
\begin{equation}
    \frac{\omega^{AFM}_{cm}(q_{\parallel}=0^{+})}{\omega^{AFM}_{bulk}(q=0)}=\sqrt{\frac{z_{\parallel}J+K}{zJ+K}}. 
\end{equation}
Since $z_{\parallel}<z$ for all cubic lattices, this confined magnon is well separated from the lowest bulk mode and should be easy to observe for simple antiferromagnets, \emph{provided that $K_s$ is lower than $K$}.

Table~\ref{TableAFM} shows our predicted frequencies for three example antiferromagnets. In the absence of measurements of $K_s$ for these materials we assumed $K_s=K-\varepsilon J$ with $\varepsilon J\ll 1$. 
\begin{figure}
	\centering
	\includegraphics[width=0.5\textwidth]{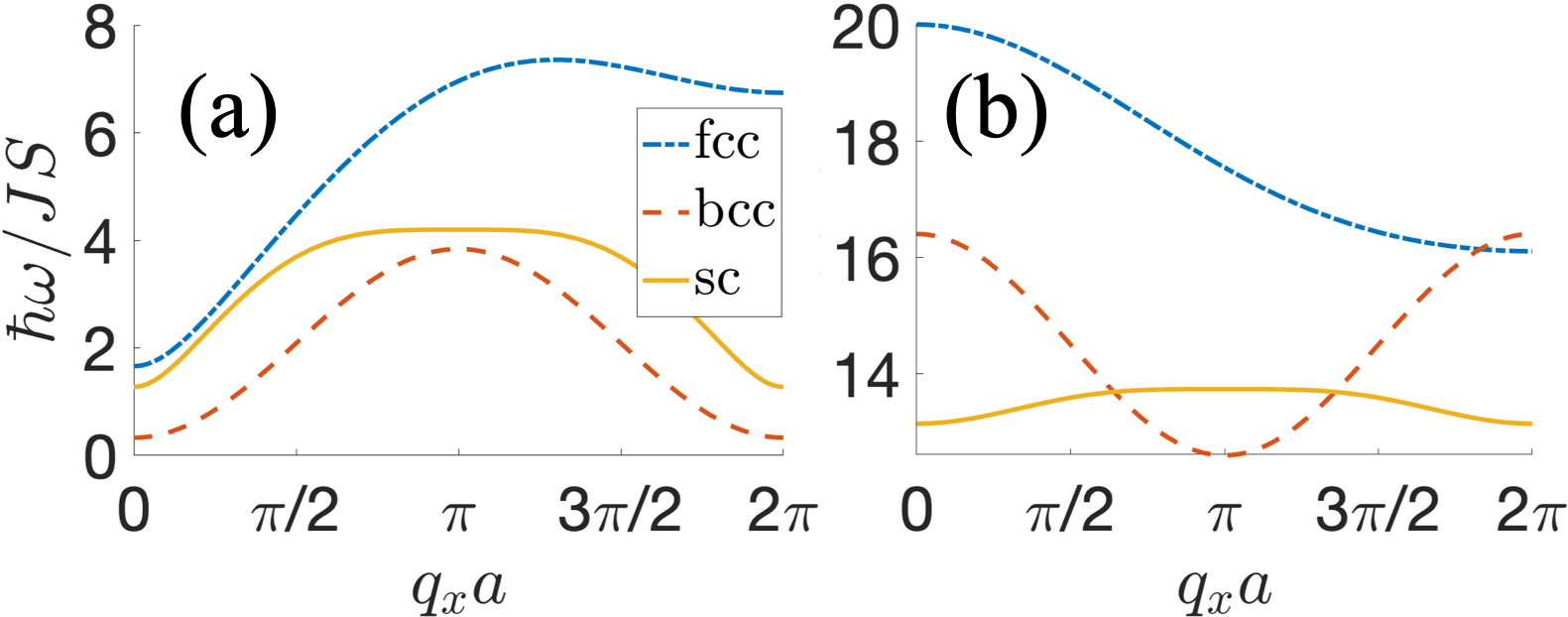}
	\caption{Dispersion for antiferromagnetic confined magnons [Eq.~(\ref{cmAFM})] for the	family of \{100\} interfaces of three different cubic lattices (face-centered, body-centered, simple cubic), and $\bm{q_{\parallel}}=(q_x,0)$. (a) $K/|J|=0.2$ and $\Delta K_s/|J|=-0.28$; (b) $K/|J|=0.2$ and $\Delta K_s/|J|=4.1$.}
	\label{AFM_cm_dispersion}
\end{figure}

Figure~\ref{AFM_cm_dispersion} shows the AFM confined magnon dispersion for the family of \{100\} interfaces in the three cubic lattices (face-centered, body-centered, simple cubic), for (a) $K/|J|=0.2$, $\Delta K_s/|J|=-0.28$, and (b) $K/|J|=0.2$, $\Delta K_s/|J| =4.1$. The behaviour is similar to FMs in that for $\Delta K_s <0$ the confined magnon is a low-frequency ``acoustic mode''; in contrast, when  $\Delta K_s >z_{\perp}/2$ it becomes instead a high-frequency ``optical mode''. The lowest optical mode occurs away from the $\bm{q_{\parallel}}=0$ zone center for both the bcc and fcc lattices. 

We showed that the confined AFM modes, like the unconfined ones, arise from to the coupling of oscillations at wavevectors $\bm{q}_{\parallel}$ and $\bm{q}_{\parallel}+\bm{Q}_{\parallel}$. The next section examines what happens in one dimension when these modes do not exist. 

\subsection{Antiferromagnetic models in one dimension: Special ``edge'' magnon\label{subsection:AFM1d}}

So far we showed that the phenomena of magnon confinement is qualitatively similar for FMs in dimensions $d=1,2,3$ and AFMs in $d=2,3$, in that their inverse length scale Eq.~(\ref{kappa_perp_final}) is identical.
We now show that the edge mode occurring in $d=1$ AFMs is qualitatively different.

Consider a spin chain with $\bm{r}_j=ja\bm{\hat{x}}$ for $j=0,1,2,\ldots$, and open boundary condition (b.c.) so that $\bm{v}_j=+a\bm{\hat{x}}$ for $j=0$ and $\bm{v}_j=\pm a\bm{\hat{x}}$ for $j\geq 1$. Plug the following trial solution into Eqs.~(\ref{AFMclassicalLEOM}):
\begin{subequations}
\begin{eqnarray}
    \delta s^{+}_{j}&=&e^{-\alpha j} \delta s^{+}_{A}\;{\rm for}\;j\;{\rm even},\\
        \delta s^{+}_{j}&=&e^{-\alpha j} \delta s^{+}_{B}\;{\rm for}\;j\;{\rm odd},
\end{eqnarray}
\end{subequations}
where $\alpha = (-iQ_{\perp}+\kappa_\perp) a$ is a complex number (we shall see that this time $Q_{\perp}$ may assume values other than $0$ and $\pi$). 
This leads to three coupled equations:
\begin{subequations}
\begin{eqnarray}
\left(\tilde{\omega}-1-\frac{2K_s}{J}\right)\delta s^{+}_{A} -e^{-\alpha}\delta s^{+}_{B}&=&0,\label{1d1}\\
\left(-\tilde{\omega}-2-\frac{2K}{J}\right)\delta s^{+}_{B} -2\cosh{(\alpha)}\delta s^{+}_{A}&=&0,\label{1d2}\\
\left(\tilde{\omega}-2-\frac{2K}{J}\right)\delta s^{+}_{A} -2\cosh{(\alpha)}\delta s^{+}_{B}&=&0.\label{1d3}
\end{eqnarray}
\end{subequations}
The last two give rise to the characteristic equation
\begin{equation}
    4\left(1+\frac{K}{J}\right)^{2}-\tilde{\omega}^{2}-4\cosh^{2}{(\alpha)}=0.
    \label{charAFM1d}
\end{equation}
Subtract Eq.~(\ref{1d3}) from Eq.~(\ref{1d1}) to get $\delta s^{+}_{B}/\delta s^{+}_{A}$, and equate to Eq.~(\ref{1d2}). This leads to \begin{equation}
    e^{-\alpha}=\frac{2\cosh{(\alpha)}\left(1-2\frac{\Delta K_s}{J}\right)^{-1}}{\tilde{\omega}+2\left(1+\frac{K}{J}\right)}.
\label{emalphaAFM1d}
\end{equation}
\RdS{Note that the fraction $2\Delta K_s/J$ is the same as the one appearing in Eq.~(\ref{kappa_perp}) for a spin chain with $z_{\perp}=2$}. Now add this equation to its inverse and the terms with $\alpha$ cancel out, leading to a quadratic equation for $\tilde{\omega}$. The candidate confined magnon is the positive root with frequency given by
\begin{eqnarray}
\frac{\hbar\omega^{AFM}_{cm, 1d}}{JS}= \frac{1}{2}\left[\left(1-2\frac{\Delta K_s}{J}\right)^{-1}-\left(1-2\frac{\Delta K_s}{J}\right)\right]\nonumber\\
+ \frac{2}{J}\sqrt{\left[\frac{\Delta K_{s}^{2}}{J-2\Delta K_s}-K\right]\left[\frac{\Delta K_{s}^{2}}{J-2\Delta K_s}-K-J\right]}. 
\label{cmAFM1d}
\end{eqnarray}
For this to correspond to a valid confined magnon solution, it must lead to $\left|e^{-\alpha}\right|<1$ or $\Re{(\alpha)=\kappa_{\perp}a>0}$. Plug $\cosh{(\alpha)}$ from Eq.~(\ref{charAFM1d}) into Eq.~(\ref{emalphaAFM1d}) to get 
\begin{equation}
    e^{-\alpha}=\frac{1}{\left(1-2\frac{\Delta K_s}{J}\right)}\sqrt{\frac{2(1+\frac{K}{J})-\tilde{\omega}}{2\left(1+\frac{K}{J}\right)+\tilde{\omega}}},
\label{emalpha}
\end{equation}
and plug Eq.(\ref{cmAFM1d}) to check whether $\left|e^{-\alpha}\right|<1$.
With numerical calculations we find that 
Eq.~(\ref{cmAFM1d}) is indeed a confined magnon when $\Delta K_c \leq \Delta K_s\leq \Delta K_{c2}$, and $\Delta K_s \geq J/2$. Here $\Delta K_c$ is identical to the result obtained for FM and other AFM cases (Eq.~(\ref{DeltaKc})), but now a new critical value $0<\Delta K_{c2}<J/2$ appears. We could not determine the value of $\Delta K_{c2}$ analytically but numerical calculations indicate that it increases with increasing $K$. 

Apart from the $\Delta K_c$ this confined ``edge'' magnon is quite different from the FM and AFM cases described previously. To see this, solve the $\Delta K_s=0$ explicitly: Eq.~(\ref{cmAFM1d}) leads to 
$\tilde{\omega}_{cm}=2\sqrt{(K/J)(1+K/J)}$, and Eq.~(\ref{emalpha}) to
\begin{equation}
    \kappa_{\perp}=\frac{1}{2a}\ln{\left(\frac{\sqrt{J+K}+\sqrt{K}}{\sqrt{J+K}-\sqrt{K}}\right)}.
\label{kappaperp1d}
\end{equation}
These should be compared to the results of section~\ref{AFM2d3d} with 
parameters appropriate for a spin chain, $z_{\perp}=2,z_{\parallel}=0,q_{\parallel}=0$. According to Eqs.~(\ref{kappa_perp_final2})~and~(\ref{cmAFM}) we would get $\tilde{\omega}_{cm}=2K/J$ and $\kappa_{perp}=0$, i.e. no confined magnon exists for $\Delta K_s=0$. In contrast,  Eq.~(\ref{kappaperp1d}) shows that in fact the confined ``edge'' magnon does exist for $\Delta K_s=0$ and $K>0$ with 
\begin{equation}
    \frac{\omega^{AFM}_{cm,1d}}{\omega^{AFM}_{bulk,1d}(q=0)}=\sqrt{\frac{J+K}{2J+K}}<1.
\end{equation}
This shows that the $\Delta K_s=0$ confined magnon is well separated from the bulk magnons, even when $K/J$ is quite small. 

This confined excitation was not apparent in previous studies in finite spin chains, because only quite low values $K/J=0.001$ were considered, making the confined magnon length scale (\ref{kappaperp1d}) comparable to the system size.\cite{Wieser2008}

\section{General theory for confined magnons in finite systems and their spin-scattering function \label{section:general}}

We now describe a general theory applicable to finite systems with arbitrary model Hamiltonian.
Our goal is to evaluate the spin-scattering function,
\begin{equation}
S_{\alpha \beta}(\bm{q},\omega) = \frac{1}{2 \pi N} \sum_{j,k} \int dt \textrm{e}^{- i \omega t}\textrm{e}^{-i  \bm{q} \cdot ( \bm{r}_j-\bm{r}_k)}\expval{s_{j}^{ \alpha}(0)s_{k}^{\beta}(t)}_T,
\label{eq:scat1}
\end{equation}
and access the observability of the confined magnon excitations predicted in Section~\ref{section:analytic} above. 

Here $N$ is the number of magnetic ions (spins), $s_{j}^{ \alpha}(t)$ denotes the $\alpha=x',y',z'$ component of the Heisenberg representation for the dimensionless spin operator describing the magnetic ion located at position $\bm{r}_j$, and $\langle \cdot\rangle_{T}$ denotes a quantum and thermal average at temperature $T$. Defined this way, Eq.~(\ref{eq:scat1}) displays resonances when $\omega$ and $\bm{q}$ match the dispersion relation for magnon propagation, along with much more information on non-dispersive (confined) modes and off-resonant excitations. 

The spin scattering function (\ref{eq:scat1}) at $q\approx \omega/c \approx 0$ (with $c$ the speed of light) also describes the spectral weight for inelastic spin excitations that satisfy the energy and momentum conservation constraints characteristic of all \RdS{photon}~scattering experiments. Different experiments (\RdS{X-ray}, Raman, THz spectroscopy) have additional selection rules that are usually accounted for using symmetry-based approaches.\cite{Cottam1986} Therefore, we can interpret  $S_{\alpha,\beta}(q=0,w)$ as a proxy for the strength of \RdS{photon}~resonances that \emph{can} occur, but one should keep in mind that which resonances get activated depend on the type of experiment and underlying symmetry of the material. 

In contrast to Section~\ref{section:analytic}, evaluating Eq.~(\ref{eq:scat1}) requires a full quantum approach based on the Holstein-Primakoff representation.\cite{Holstein1940} The method we present here is an adaptation to non-translation invariant systems of the framework for evaluating the scattering function presented in \onlinecite{Fishman2018}.

\subsection{Formal diagonalization and magnon frequencies\label{subsec:formaldiag}}

We start by using the Holstein-Primakoff transformation \cite{Holstein1940} to represent spin operators as Bosonic creation and destruction operators, $a_{j}^{\dag}$ and $a_{j}$, respectively, where $j$ again labels the lattice site. 
For spins with quantum number $S$ we get
\begin{subequations}
\begin{eqnarray}
s^{x'}_{j}  &=& \frac{ \sqrt{2S - a^{\dag}_{j} a_{j}} \: a_{j} + a^{\dag}_{j} \sqrt{2S -a^{\dag}_{j} a_{j}}}{2} ,\label{eq:HPx}\\
s^{y'}_{j} &=& \frac{ \sqrt{2S - a^{\dag}_{j} a_{j}} \: a_{j} - a^{\dag}_{j} \sqrt{2S -a^{\dag}_{j} a_{j}}}{2j},\label{eq:HPy}\\
s^{z'}_{j} &=&  (S-a^{\dag}_{j} a_{j}).\label{eq:HPz}
\end{eqnarray}
\label{eq:HP}
\end{subequations}
It is easy to check that the Bosonic commutation relation $[a_{i},a_{j}^{\dag}]=\delta_{ij}$ implies $[ s^{\alpha}_{i}, s^{\beta}_{j} ] = i\epsilon_{\alpha \beta \gamma} s^{\gamma}_{i}\delta_{ij}$ for the spin operators. The $s^{z'}_{j}$ eigenstates $\ket{S,m_j}$ with $m_j=-S,-S+1,\ldots,S$ are relabeled as $\ket{m'_{j}}$ with $m'_{j}=S-m_j=0,1,\ldots,2S$ denoting the number of ``spin-flip'' excitations in each site. They satisfy 
$a_{j} \ket{m'_j} = \sqrt{m'_j}\ket{m'_j-1}$ for $0 \leq m'_j \leq 2S$, and $a^{\dag}_{j} \ket{m'_j} = \sqrt{m'_j+1} \ket{m'_j+1}$ for $0 \leq m'_j \leq 2S - 1$. Note that $m_j=0$ corresponds to the ``vacuum'' of Holstein-Primakoff excitations, which possesses the maximum spin. In a simple FM model this will be the ground state; in contrast, for AFM models we will have to define a set of Holstein-Primakoff operators for each sublattice of the system, so that the vaccum state is the maximum spin state $m_j=S$ in one sublattice together with the minimum spin state $m_j=-S$ in the other. 
From here, we limit ourselves to a small number of excitations $m'_j\ll 2S$, which is always a good approximation at low $T$ and large $S\gg 1$. This allows us to approximate Eqs.~(\ref{eq:HPx})~and~(\ref{eq:HPy}) as
$s^{+}_{j}=s^{x'}_{j}+is^{y'}_{j} \approx  \sqrt{2S} a_j$, and $s^{-}_{j}=s^{x'}_{j}-is^{y'}_{j} \approx  \sqrt{2S}a^{\dag}_{j}$. 

Plugging Eqs.~(\ref{eq:HPx})--(\ref{eq:HPz}) into the interacting spin Hamiltonian ${\cal H}$ leads to three contributions that scale as different powers of spin $S$,
\begin{equation}
 {\cal H}  =  {\cal H}_0(S^2) +  {\cal H}_1(S^1) +  {\cal H}_2 (S^0).
 \label{eq:expand}
\end{equation} 
Here ${\cal H }_0 $ is the ground state energy of the system, a constant proportional to $S^2$ which does not contain any $a_j$ or $a_{j}^{\dag}$ terms, so it can be dropped. 
The next contribution ${\cal H }_1$ is linear in $S$, and quadratic in $a_j$ and $a_{j}^{\dag}$;
this is the magnon Hamiltonian. The last contribution ${\cal H}_2$ is independent of $S$, and contains quartic and higher order terms such as $a_{i}^{\dag} a_j a_{k}^{\dag} a_l$. It describes the mutual interaction between magnons in the system. ${\cal H}_{2}$ does not play a role at low $T$ when the number of thermally activated magnons is small. For this reason, we are not considering ${\cal H}_2$ and focus entirely on ${\cal H}_1$. We note that if we divide Eq. (\ref{eq:expand}) by $S^2$, it becomes an expansion in powers of $1/S$. Therefore, keeping ${\cal H}_1$ and neglecting ${\cal H}_2$ is equivalent to keeping a $1/S$ contribution and dropping a $1/S^2$ correction. Evidently this becomes a good approximation in the limit $S\gg 1$.

In all cases we can write ${\cal H}_1$ in matrix form, 
\begin{equation}
{\cal H}_1  = \bm{v}^{\dag}\cdot \bm{L} \cdot\bm{v},
\end{equation}
where $\bm{v}^{\dag} = \left( a_{1}^{\dag}, a_{2}^{\dag}, \hdots, a_{N}^{\dag} | a_{1}, a_{2}, \hdots, a_{N} \right)$, and $\bm{L}$ is an Hermitian $2N \times 2N$ matrix with the following block structure:
\begin{equation}
 \bm{L} = \begin{bmatrix}  \bm{P} & \bm{Q} \\ \bm{Q}^* & \bm{P}^* \end{bmatrix},
 \label{eq:LMatrix}
 \end{equation}
 where $\bm{P}=\bm{P}^{\dag}$ and $\bm{Q}=\bm{Q}^{T}$ are $N \times N$ matrices.

It is now time to make our first deviation from the conventional bulk approach.\cite{Fishman2018} In the bulk approach, at this point, we would apply a space Fourier transform to $\bm{v}$, so that $\bm{L}$ is reduced to a $2 u \times 2 u$ matrix where $u$ is the number of sites in the magnetic unit cell. For a finite system with open boundary condition (b.c.) this approach is no longer useful and we instead diagonalize $\bm{L}$ numerically. 
The structure of Eq.~(\ref{eq:LMatrix}) is due to particle-hole symmetry and implies the set of eigenvalues:
\begin{equation}
\epsilon_n   =
\left\{
	\begin{array}{ll}
		\hbar \omega_n /2  & \mbox{if } n \in [0,N-1) \\
		\hbar \omega_{n-N}/2 & \mbox{if } n \in [N,2N-1].
	\end{array}
\right. 
\end{equation}
We diagonalize $ \bm{L} $ using unitary transformation $ \bm{L} ' = \bm{U}  \bm{L}  \bm{U}^\dag$, with the columns of $\bm{U}^\dag$ being the eigenvectors of $\bm{L}$. We diagonalize ${\cal H}_1$ by transforming it via,
\begin{equation}
{\cal H}_1 =  \bm{v}^{\dag} \bm{U}^{\dag}  \bm{U} \bm{L}  \bm{U}^{\dag}  \bm{U} \bm{v} = \bm{w}^{\dag} \bm{L}' \bm{w},
\end{equation}
where 
\begin{equation}
    \bm{w}^{\dag} =  \bm{v}^{\dag}  \bm{U}^{\dag} = \left( \alpha_{0}^{\dag}, \alpha_{1}^{\dag}, \hdots, \alpha_{N-1}^{\dag} | \alpha_{0}, \alpha_{1}, \hdots, \alpha_{N-1} \right)
\end{equation}
is the vector of bosonic operators that describe the normal modes of the system (a new set of creation/annihilation operators satisfying $[\alpha_n,\alpha^{\dag}_{n'}] = \delta_{n,n'}$). 
Expanding ${\cal H}_1 $ on this new basis, we get
\begin{equation}
{\cal H}_1 = \sum_{n = 0}^{N-1} \frac{\hbar \omega_n}{2} \left( \alpha_{n}^{\dag} \alpha_{n} + \alpha_{n} \alpha_{n}^{\dag} \right) = \sum_{n=0}^{N-1} \hbar\omega_{n}(\alpha_{n}^{\dag} \alpha_{n} + \frac{1}{2}).
\end{equation}
Therefore, $\alpha_{n}^{\dag}$ and $\alpha_n$ create and annihilate oscillating collective excitations with energy $\hbar \omega_n$, i.e., they describe magnon excitations. 
In the presence of periodic b.c., these excitations are superpositions of forward and backward propagating waves, leading to sinusoidal (standing) spin waves. 
As we shall see in the solution for the FM and AFM chains with open b.c., the standing modes become anharmonic. 

\subsection{Calculation of the Scattering Function}

Our method allows the evaluation of Eq.~(\ref{eq:scat1}), $S_{\alpha \beta}(\bm{q},\omega)$, for any $\alpha$ and $\beta$. However, without adding simplifying assumptions to the system, we must repeat the process for each unique combination of $\alpha$ and $\beta$. Therefore, we limit ourselves to collinear systems such as a Heisenberg FM or AFM with single ion anisotropy where the spins are aligned along $\pm \bm{z}'$, and the total spin along $\bm{z}'$ is a constant of motion. This restriction means $S_{\alpha \beta}(\bm{q},\omega) = 0$ for $\alpha \neq \beta$, $S_{z'z'}(\bm{q},\omega) = 0$, and $S_{x'x'}(\bm{q},\omega) =S_{y'y'}(\bm{q},\omega)\neq 0$. Below we focus on $S_{x'x'}(\bm{q},\omega)$ as it gives a complete description of collinear systems.

From $\bm{v} = \bm{U}^\dag \bm{w}$ we get
\begin{subequations}
\begin{eqnarray}
		a_j   &=& \sum_{n=0}^{N-1} \left( U_{j,n}^\dag \alpha_n + U^\dag_{j,n+N} \alpha_{n}^{\dag} \right),\\
		a_{j}^{\dag} &=& \sum_{n=0}^{N-1} \left( U_{j+N,n}^{\dag} \alpha_n + U^{\dag}_{j+N,n+N} \alpha_n^{\dag} \right),
\end{eqnarray}
\end{subequations}
which allows expressing $s^{x'}_{j}(t)$ in terms of normal mode operators $\alpha_n(t)=\textrm{e}^{-i\omega_{n} t} \alpha_{n}$,
\begin{eqnarray}
  s_{j}^{x'}(t)  &\approx&  \frac{\sqrt{2S}}{2}[a_j(t) + a_j^{\dag}(t)]\nonumber\\
  &=& \sqrt{\frac{S}{2}} \sum_{n=0}^{N-1} \left[ \textrm{e}^{-i\omega_{n} t} \alpha_{n} X_{j,n} + {\rm H.c.}
  \right], \label{eq:sjxt}
 \end{eqnarray}
where 
\begin{subequations}
\begin{equation}
X_{j,n}=U_{j,n}^\dag+U_{j+N,n}^\dag.
\end{equation}
\end{subequations}
Pluging Eq.~(\ref{eq:sjxt}) into the spin-spin correlation function and noting that $\expval{\alpha_{n}^{\dag}\alpha^{\dag}_{n'}}_T=\expval{\alpha_{n}\alpha_{n'}}_T=0$, and
$\expval{\alpha_{n}^{\dag}\alpha_{n'}}_T=n_B(\omega_n)\delta_{nn'}$, 
where $n_B(\omega_n)$ is the Bose function,
\begin{equation}
n_B(\omega_n) = \frac{1}{\textrm{e}^{\hbar \omega_n / k_B T} -1},
\label{eq:bo}
\end{equation}
we get 
\begin{eqnarray}
  \expval{s_{i}^{x'}(0)s_{j}^{x'}(t)}_T &=&  \frac{S}{2} \sum_{n=0}^{N-1}  \left\{X_{i,n}^{*}X_{j,n} n_B(\omega_n) \textrm{e}^{-i \omega_n t} \right.\nonumber\\
  &&\left.+X_{i,n}X_{j,n}^{*}[n_B(\omega_n)+1] \textrm{e}^{i \omega_n t}\right\}.
\end{eqnarray}
Finally, taking the Fourier transform and noting that $\delta(\omega+\omega_n)$ is always zero for $\omega>0$ we get our explicit expression for the spin-scattering function,
\begin{eqnarray}
S_{x'x'}(\bm{q},\omega) &=& \frac{S}{2 N} \sum_{n=0}^{N-1} 
\left|\sum_{j}
\textrm{e}^{-i\bm{q} \cdot \bm{r}_j} X_{j,n}
\right|^{2}
[n_B(\omega_n)+1] \nonumber\\
&&\times\delta(\omega - \omega_n).
\label{eq:ScatF}
\end{eqnarray}

Below we replace the delta functions $\delta(\omega-\omega_n)$ by a smooth Gaussian with broadening set by the energy resolution of the measurements involved:
\beq
\delta(\omega-\omega_n) = \frac{1}{\sqrt{2\pi \Delta^2}}\textrm{e}^{-\frac{(\omega-\omega_n)^{2}}{2\Delta^2}}. 
\eeq
For definiteness we use $\Delta=0.02 |J|S$ for our numerical calculations.

\section{Application to one-dimensional models\label{section:applicatonto1d}}

In $d=1$  Hamiltonian (\ref{Hinfty}) becomes
\begin{eqnarray}
{\cal H} &=& J \sum_{j=0}^{N-2} \bm{s}_j \cdot \bm{s}_{j+1}   +b J \bm{s}_{N-1}\cdot \bm{s}_0\nonumber\\
&&- K \sum_{j=1}^{N-2}(s_{j}^{z'})^{2}-K_{s0}\left(s_{0}^{z'}\right)^{2} -K_{sN-1} \left(s_{N-1}^{z'}\right)^{2},
\label{H}
\end{eqnarray}
where parameter $b$ describes the choice of b.c.. We set either $b=1$ for periodic b.c. or $b=0$ for open b.c.. 
Our results below are independent of the relative orientation between easy axis $\bm{\hat{z}}'$ and the spin chain direction, which can be taken as arbitrary.

\subsection{Ferromagnet}

For $J<0$ and uniform FM ground state 
the matrix $\bm{L}$ of Eq.~(\ref{eq:LMatrix}) is given by
\begin{widetext}
\begin{equation}
\begin{split}
\bm{P} = \frac{1}{2}|J|S
\begin{bmatrix}
1+ b+ \frac{2K_{s0}}{|J|} &- 1 &  0 & 0 & \hdots & -b \\
-1 & 2\left(1 + \frac{K}{|J|}\right)& -1 &  0 & \hdots & 0 \\
0 & -1 & 2\left(1 + \frac{K}{|J|}\right)& -1 & \hdots & 0 \\
\vdots & \vdots &\vdots & \ddots &  \hdots & \vdots \\
0 & 0 & 0 & -1 & 2\left(1 + \frac{K}{|J|}\right)& -1\\
-b & 0 & 0 & 0 & -1 & 1+ b+ \frac{2K_{sN-1}}{|J|}
\end{bmatrix},
\text{ and }
\bm{Q} =\bm{0}.
\end{split}
\end{equation}
\end{widetext}
We diagonalize this matrix numerically, and use the resulting eigenvalues and eigenvectors to compute the scattering function explicitly using Eq.~(\ref{eq:ScatF}). 

\begin{figure}
	\centering
	\includegraphics[width=0.5\textwidth]{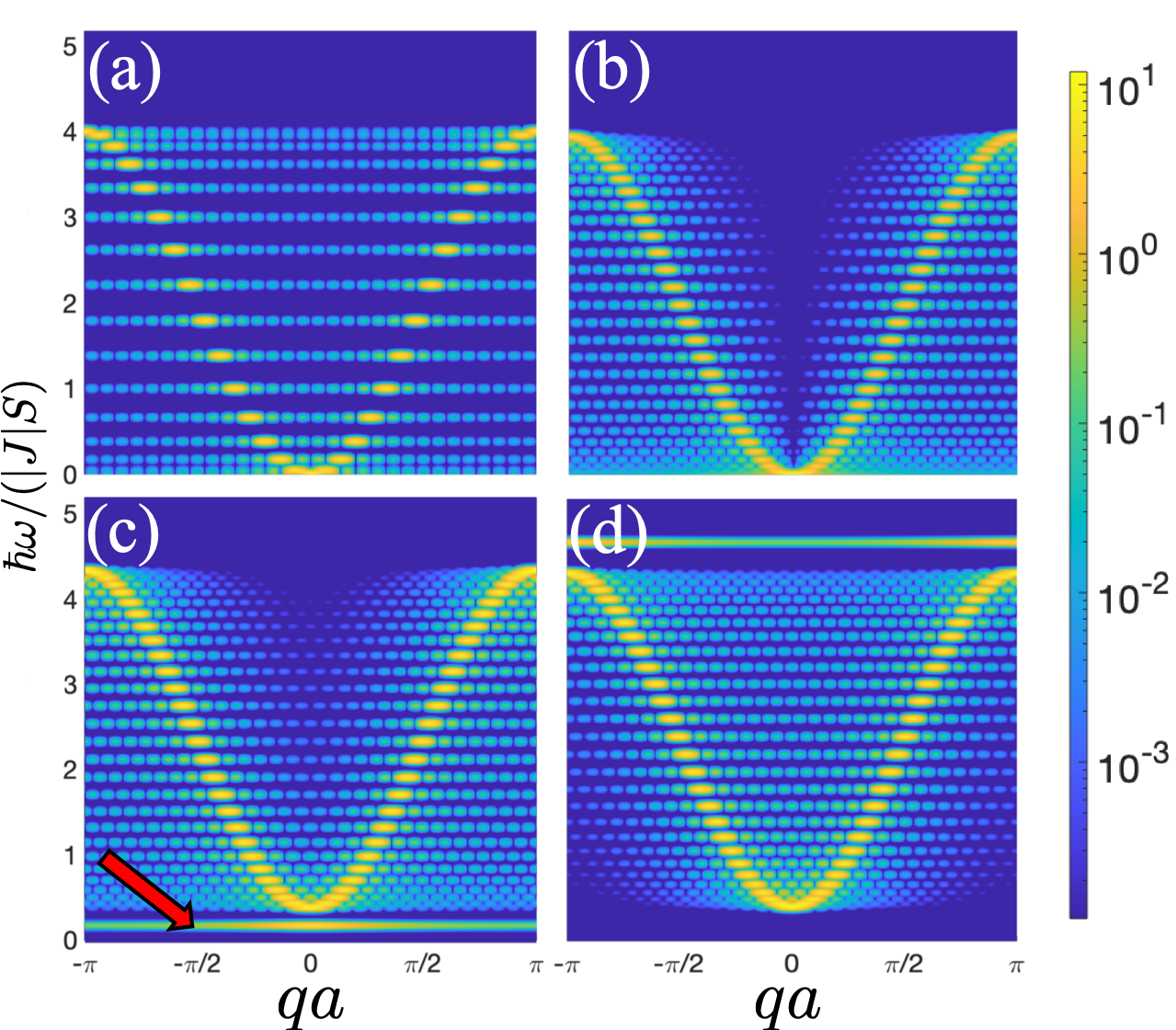}
	\caption{Spin scattering function $S_{x'x'}(\omega, \bm{q})$ [Eq.~(\ref{eq:ScatF})] for the FM model with $N=30$ spins. (a) $K=K_s=0$ and periodic b.c.; (b) $K=K_s=0$ and open b.c.; (c) $K/|J|=0.2, \Delta K_s/|J|=-0.28$, open b.c.; (d) $K/|J|=0.2, \Delta K_s/|J|=1.4$, open b.c.. The plot is a logarithmic heat map of $S_{x'x'}(\omega, \bm{q})$ as a function of frequency $\omega$ and wavevector $\bm{q}$ along the spin chain direction.  In all cases $S_{x'x'}(\bm{q}, \omega)$ has resonances at the usual FM magnon dispersion relation. The granularity of the heat map reflects quantization of the spin wave frequencies and quasimomenta. In (c) the red arrow points to an acoustic resonance (the confined magnon) at low frequencies. In (d) we chose a quite high value of edge anisotropy $\Delta K_s/|J|=1.4$ to show that the confined magnon becomes a high frequency ``optical'' resonance.}
	\label{FM_heatmaps}
\end{figure}

\begin{figure}
	\centering
	\includegraphics[width=0.5\textwidth]{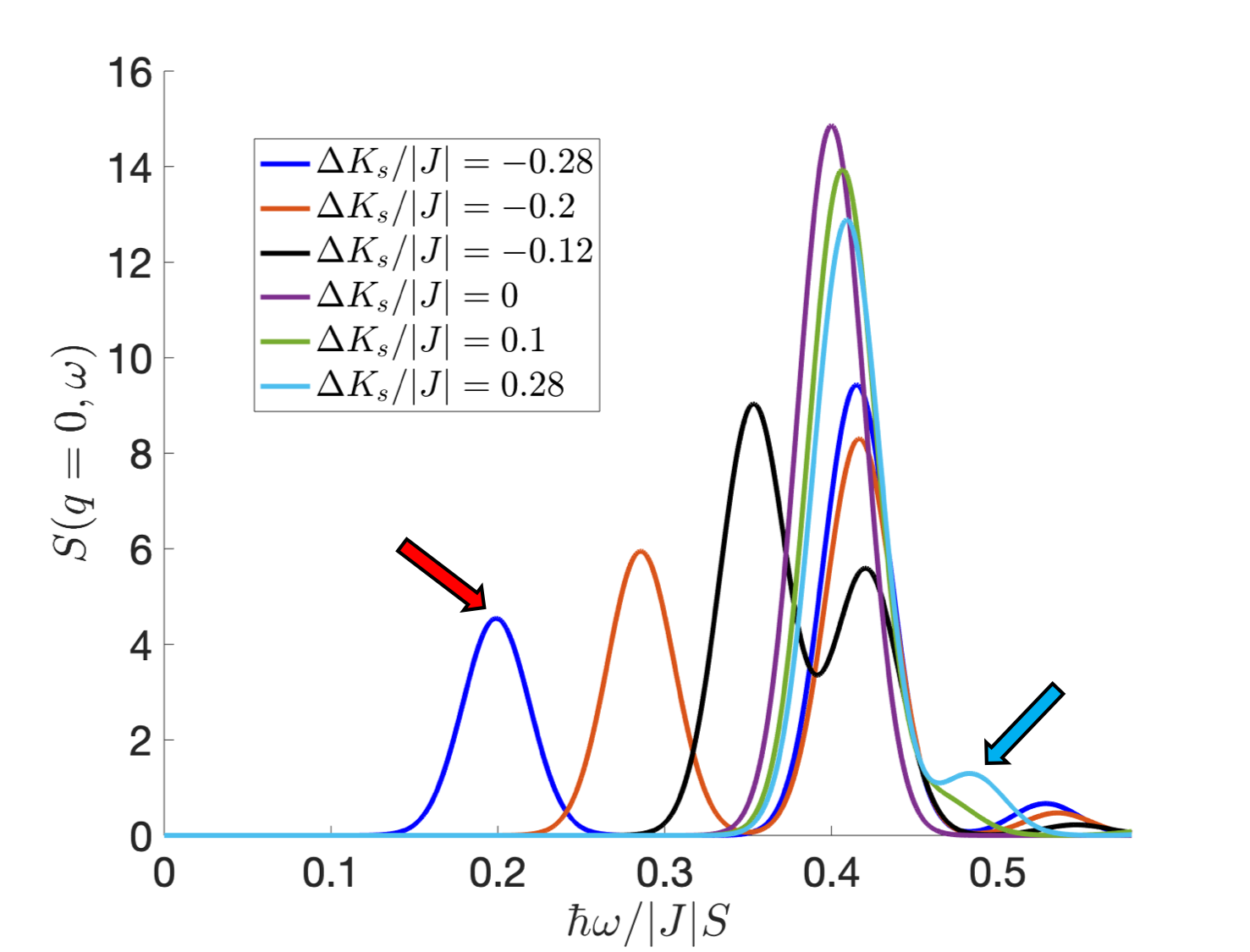}
	\caption{Spin scattering function $S_{x'x'}(q=0,\omega)$ (proxy for optical experiments) of the $d=1$ FM model for $K/|J|=0.2$ and various $\Delta K_s$. The large peak at $\tilde{w}\approx 0.4$ is the bulk $q=0$ magnon; smaller resonances (see arrows) are acoustic confined magnons.}
	\label{fig:FMacousticShoulder}
\end{figure}

Figures~\ref{FM_heatmaps}(a, b) shows the $\omega$ vs. $q$ heatmap for the spin scattering function for (a) periodic and (b) open b.c., and $K=K_s=0$ (no magnetic anisotropy). In both cases $S_{x'x'}(\bm{q},\omega)$ is sharply peaked at the $d=1$ bulk FM dispersion Eq.~(\ref{bulkFM}),
\beq
\frac{\hbar\omega^{FM}_{bulk}(\bm{q})}{|J|S}=\frac{2K}{|J|}+ 4\sin^{2}{\left(\frac{qa}{2}\right)}, 
\eeq
with no noticeable modifications to the dispersion due to open b.c. (Note the logarithmic scale for the color code). 

The granularity seen in the heat maps is a consequence of quantization of magnon frequencies and quasimomenta in a finite system. A noticeable difference is that frequency quantization in Fig.~\ref{FM_heatmaps}(a) (periodic b.c.) is twice as large as in Fig.~\ref{FM_heatmaps}(b) (open b.c.). This occurs because for periodic b.c. the modes are two-fold degenerate. These can be chosen to have $s^{x'}_{j}(t)\propto \sin{(Q r_{j})}$ or $\cos{(Q r_{j})}$ for a single wavevector $Q$. 
Hence, for periodic b.c. all modes with $Q\neq 0,\pi/a$ are two-fold degenerate. In contrast, under open b.c. these modes \RdS{become anharmonic (see below)}~and are no longer degenerate because they cause different fluctuations at the edges. 

In finite systems with periodic b.c. the spin-spin correlation is sinusoidal, i.e. $s_{j}^{x'}(0)s_{k}^{x'}(t)$ is a linear combination of 
$\textrm{e}^{iQa(j-k)}$ and $\textrm{e}^{-iQa(j-k)}$ with wavevector $Q$ assuming one of the $N$ special points inside the first Brillouin zone ($Q=2\pi m_Q/N$ for integers $m_Q=-N/2+1,\ldots, N/2$). In this case Eq.~(\ref{eq:scat1}) is a linear combination of
\beq
\frac{1}{N}\sum_{j,k}\textrm{e}^{-i(q-Q)a(j-k)}=\frac{1}{N}
\frac{\sin^{2}{\left[\frac{(q-Q)aN}{2}\right]}}
{\sin^{2}{\left[\frac{(q-Q)a}{2}\right]}}, \label{eq:Ssin}
\eeq
and the same expression for $Q\rightarrow -Q$.
This function is maximum ($\propto N$) when $q=Q + 2\pi n$ for arbitrary integer $n$, and is exactly equal to zero when 
$q$ is a special point that is different than $Q$ (i.e., $q=2\pi m/N$ with $m\neq m_Q$). However,  Eq.~(\ref{eq:Ssin}) is nonzero when $q$ falls outside one of the $N$ special points. It is this effect that gives rise to several nonresonant peaks appearing for arbitrary $q$ in Fig.~\ref{FM_heatmaps}. 

When $N\rightarrow \infty$, Eq.~(\ref{eq:Ssin}) becomes proportional to $\delta(q-Q)$, because the amplitude of the nonresonant peaks are negligibly small in comparison to the resonances that occur when $q,\omega$ matches the FM dispersion relation. The presence of several weaker resonances away from the magnon dispersion relation is a distinctive feature of finite systems.\cite{Hendriksen1993}

For periodic b.c. the weak resonances are equally spaced along the $q$ axis, because each mode is characterized by a single wavevector $Q$. This is in contrast to the open b.c. case where we see the weak resonances expelled from the $q\approx 0$ region, signalling mode anharmonicity (i.e., each mode is no longer characterized by a single $Q$). 
Similar plots for larger $N$ (not shown) are identical to Fig.~\ref{FM_heatmaps}(a, b) but with a larger density of grains. It should be emphasized that for the case of open b.c., $q$ cannot be strictly be interpreted as momentum.  It takes the role of a bookkeeping parameter, that is only interpreted as momentum when $N\rightarrow \infty$.

When $K=K_s> 0$, a gap opens at low frequencies. With $\Delta K_s=0$ the FM model shows no noticeable difference for the periodic and open b.c. cases, apart from the shifted weak resonances. This is in agreement with Section~\ref{subsection:AnalyticFM} in that no confined magnon exists for $\Delta K_s=0$ and $q_{\parallel}=0$. 

However, when $\Delta K_s<0$ the edge spins at $j=0,N-1$ are softened, leading to the formation of a confined magnon at the edges. 
Figure~\ref{FM_heatmaps}(c) shows the case $K/|J|=0.2$ with $\Delta K_s/|J|=-0.28$: The confined magnon resonance (shown by a red arrow) is at $\tilde{\omega}_{cm}=0.20$, in the middle of the anisotropy gap -- this is the \emph{acoustic confined magnon} and its frequency is in close agreement with Eq.~(\ref{cmFM}) for $z_{\parallel}=0,z_{\perp}=2$ and $q_{\parallel}=0$. In contrast, Fig.~\ref{FM_heatmaps}(d) shows what happens when the edge spins are hardened by a large easy-axis surface anisotropy $K/|J|=0.2, \Delta K_s/|J|=1.4$: This induces the formation of an \emph{optical confined magnon} at $\tilde{\omega}_{cm}=4.76$, again in close agreement with Eq.~(\ref{cmFM}). For the optical confined magnon to be visible in spectroscopy its frequency has to be above the bulk zone-edge magnon at $q=\pi/a$. We find that this only happens for quite high $\Delta K_s$ \RdS{as shown in Fig.~\ref{FM_heatmaps}(d)}. 

Figure~\ref{fig:FMacousticShoulder} plots the proxy for \RdS{photon}~spectroscopy $S_{x'x'}(q=0,\omega)$ as a function of $\omega$, for $K/|J|=0.2$ and various $\Delta K_s$. At $\Delta K_s=0$ we see only the main bulk peak at $\tilde{\omega}\approx 0.4$; however, as $\Delta K_s$ becomes negative the confined magnon peak is formed, taking spectral weight out of the bulk peak. For $\Delta K_s \gtrsim 0$ the confined magnon is a small shoulder next to the bulk peak (blue arrow).

\subsection{Antiferromagnet}

Model Hamiltonian~(\ref{H}) with $J>0$ and 
homogeneous AFM state (\ref{gtype}) requires a minor change to the definition of the magnon operators:
$a_j,a^{\dag}_j$ for even $j$ are defined in the same way as for the FM. For odd $j$ their definition is changed to $s^{z'}_{j} = (-S+a^{\dag}_{j} a_{j})$ and $s^{+}_j=\sqrt{2S}a^{\dag}_j$. 
The $\bm{L}$ matrix 
is now given by
\begin{widetext}
\begin{equation}
\begin{split}
\bm{P} =  \frac{JS}{2}
\begin{bmatrix}
1+ b+ \frac{2K_{s0}}{J}&  0   &   0 & \cdots  & 0         \\     
0 & 2\left(1+ \frac{K}{J}\right)  & 0 & \cdots &  0        \\    
 \vdots&   &  \ddots &  &  \vdots \\ 
  0& \cdots & 0 & 2\left(1+ \frac{K}{J}\right)  & 0        \\     
 0& \cdots & 0 &0 & 1+ b+\frac{2K_{sN-1}}{J} \\   
\end{bmatrix}
\text{and }
\bm{Q} =  \frac{JS}{2}
\begin{bmatrix}
 0 & 1 & 0   &\cdots & b          \\    
1 &  0& 1 &\cdots &0          \\    
\vdots & \ddots & \ddots  & \ddots & \vdots \\ 
 0 & \cdots &1  &0  &    1    \\  
b &  0&\cdots & 1  &  0     \\  
\end{bmatrix}.
\end{split}
\end{equation}
\end{widetext}

\begin{figure}
	\centering
	\includegraphics[width=0.5\textwidth]{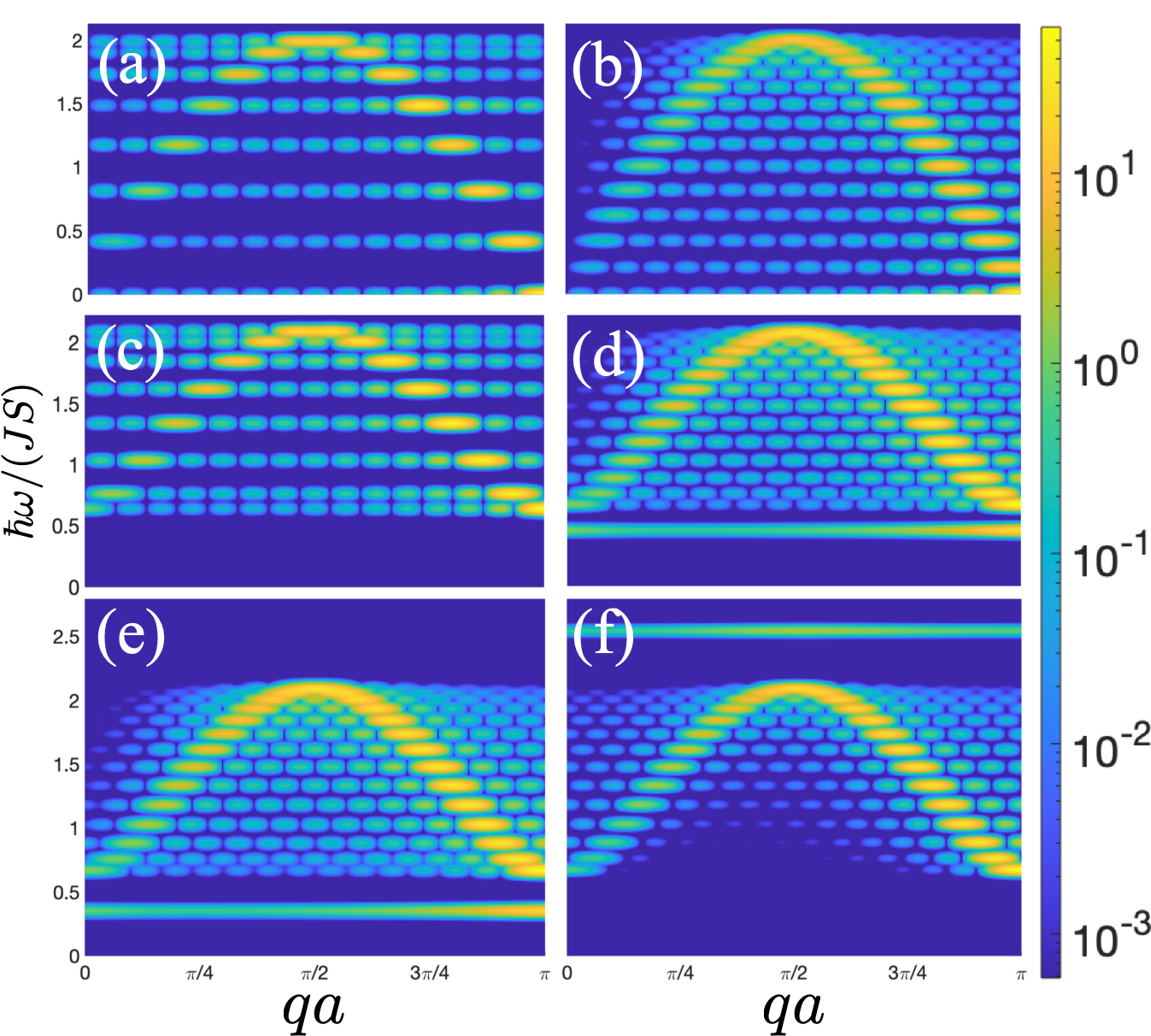}
	\caption{Spin scattering function for the AFM model with $N=30$. (a) $K=K_s=0$ and periodic b.c.; (b) $K=K_s=0$ and open b.c.; (c) $K/|J|=0.05$, $\Delta K_s=0$ and periodic b.c.; (d) $K/|J|=0.05$, $\Delta K_s=0$ and open b.c.; (e) $K/|J|=0.05$, $\Delta K_s=-0.05$ and open b.c.; (f) $K/|J|=0.05$, $\Delta K_s=0.8$ and open b.c. In all cases $S_{x'x'}(\bm{q}, \omega)$ has resonances at the usual AFM magnon dispersion relation. However, the impact of nonzero anisotropy is dramatically different. In (d)
it is seen that an acoustic confined magnon arises even when $\Delta K_s=0$ (note that this mode is absent when the b.c. is periodic, case (c)). When $\Delta K_s\neq 0$ the confined magnon frequency shifts: It decreases when the edge spin is softened (case (e) with $\Delta K_s<0$)  and it increases when it is hardened (case (f) with $\Delta K_s>0$). For large $\Delta K_s/|J|=2.8$ shown in (f) the confined magnon becomes an optical mode.}
	\label{AFM_heatmaps}
\end{figure}

Once again this is diagonalized numerically and the eigenvalues/eigenvectors are plugged into Eq.~(\ref{eq:ScatF}) to determine the spin scattering function explicitly. Figures~\ref{AFM_heatmaps}(a, b) show the $K=K_s=0$ cases with periodic and open b.c., respectively. Similar to the FM case both heatmaps show strong resonances at the $d=1$ bulk AFM magnon dispersion Eq.~(\ref{bulkAFM}),
\beq
\frac{\hbar\omega^{AFM}_{bulk}(\bm{q})}{JS}=2\sqrt{\left(1+\frac{K}{J}\right)^{2}-\cos^{2}{\left(qa\right)}},
\eeq
which is now linear in $q$ for $K=0$ and small $q$ (long wavelength). Like the FM case, the magnon frequencies are doubly degenerate for periodic b.c., and this degeneracy is lifted for open b.c.

Figures~\ref{AFM_heatmaps}(c, d) shows what happens when we turn on bulk anisotropy $K=K_s> 0$ in the periodic and open b.c. cases. This time there is a remarkable difference. While the periodic b.c. displays the usual gapped bulk spectra with no confined mode (Fig.~\ref{AFM_heatmaps}(c)), the open b.c. now has in addition a strong confined acoustic ``edge'' magnon at $\tilde{\omega}=0.46$ (Fig.~\ref{AFM_heatmaps}(d)). 
Its frequency is in close agreement with Eq.~(\ref{cmAFM1d}) for $\Delta K_s=0$.

The confined magnon energy is visibly separated from the lowest bulk mode in Fig.~\ref{AFM_heatmaps}(d) and should be visible in spectroscopy. As already mentioned in Section~\ref{subsection:AFM1d} this confined magnon was not apparent in previous studies in finite spin chains, because 
only quite low values $K/J=0.001$ were considered, making its length scale comparable to the system size.\cite{Wieser2008} 

\begin{figure}	
	\centering
	\includegraphics[width=0.48\textwidth]{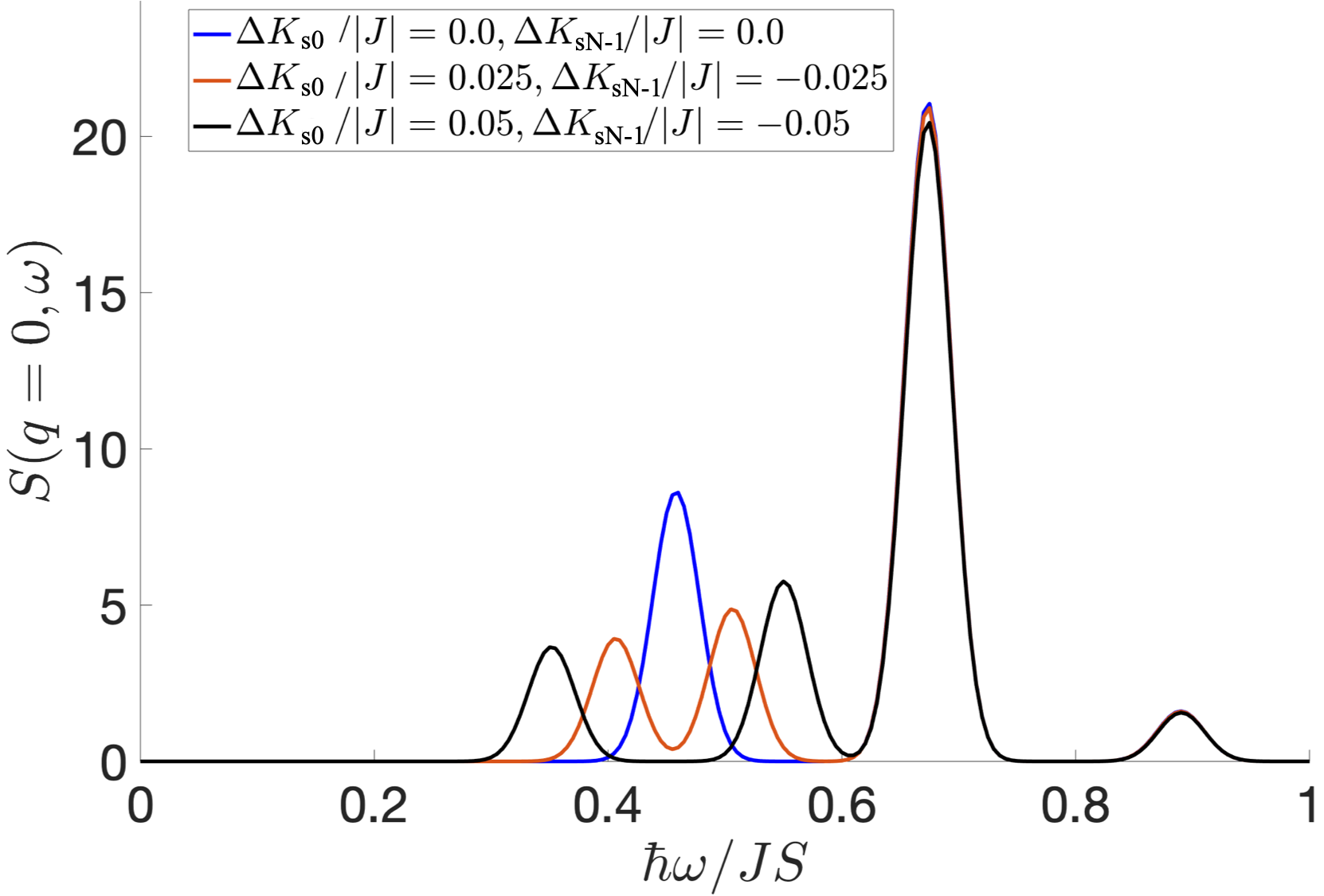}
	\caption{Spin scattering function $S_{x'x'}(q=0,\omega)$ at $q=0$ for the AFM model with $K/J=0.05$ and $\Delta K_s$ different for edges $j=0$ and $j=N-1$. The asymmetry splits the acoustic confined magnon resonance into two peaks, which are clearly visible as smaller resonances next to large bulk $q=0$ mode at $\tilde{\omega}=0.65$.}
	\label{fig:AFM_KsAsymmetric}
\end{figure}

Figure~\ref{AFM_heatmaps}(e)  shows that adding $\Delta K_s<0$ lowers the frequency of the acoustic confined magnon. This makes sense since the confined magnon is localized at one of the edges (See Fig.~\ref{fig:AFM_KsAsymmetric}(a)), and the $\Delta K_s<0$ softens the spin at both edges. 

In contrast, Fig.~\ref{AFM_heatmaps}(f) shows that a large $\Delta K_s>0$ increases the confined magnon resonance to the point that it becomes an optical mode (above the bulk zone edge magnon frequency). This is consistent with what we found for $d=2,3$ in Eq.~(\ref{cmAFM}): $\Delta K_s/|J| >z_{\perp}/2$ leads to a high-frequency ``optical'' confined magnon.

In order to illustrate the sensitivity of the acoustic confined magnon to the environment at the edges,  we set $\Delta K_s$ at spin $j=0$ different from $\Delta K_s$ at spin $j=N-1$. By making the $j=0$ edge harder and the $j=N-1$ softer, the confined magnon is split into two resonances. The lower frequency one is dominated by spin oscillations at the softer edge, while the higher frequency one contains spin oscillations concentrated at the harder edge. This situation should be quite common in nanoparticles on top of a substrate, or thin films sandwiched between two different materials. Since the confined magnon peaks are clearly separated from the bulk $q=0$ mode (large peak at $\tilde{\omega}\approx 0.65$, they are clearly detectable by \RdS{photon}~spectroscopy. 

\begin{figure}	
	\centering
	\includegraphics[width=0.5\textwidth]{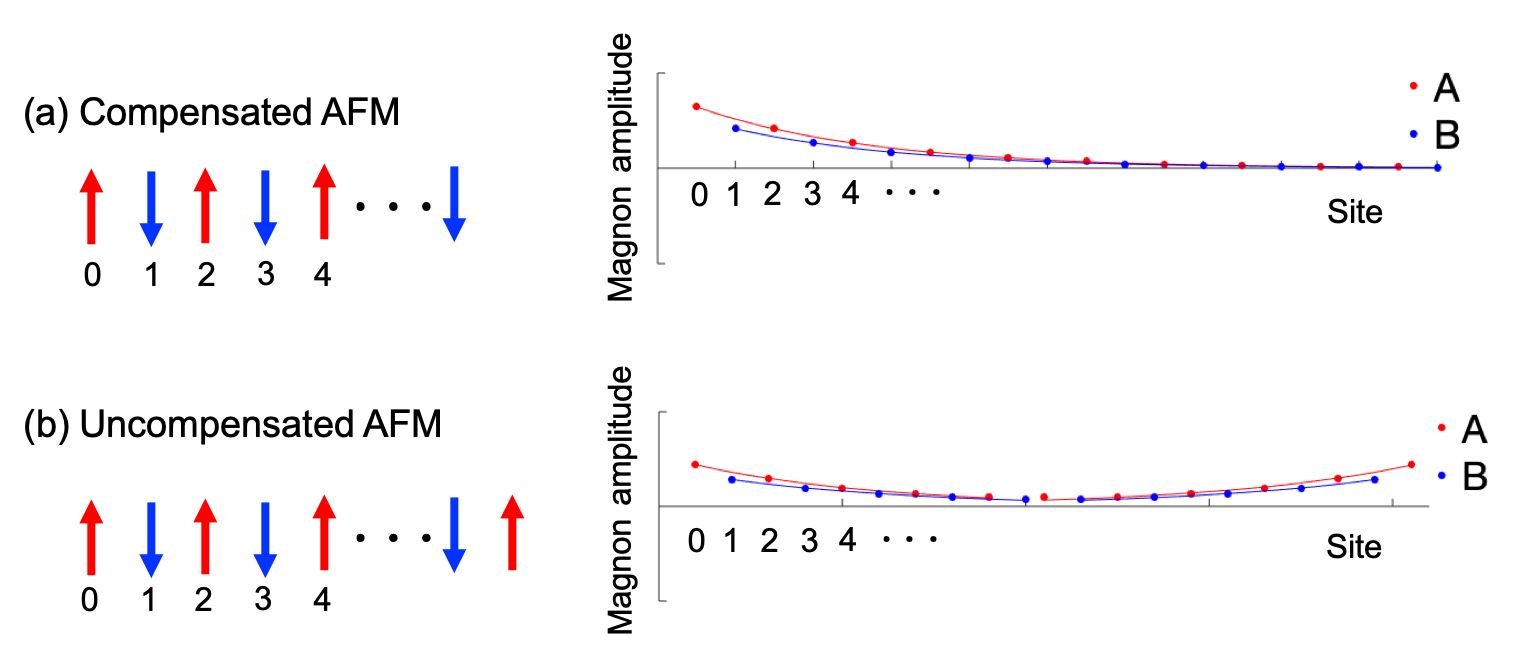}
	\caption{\RdS{Visualizing the acoustic confined magnon for (a) compensated and (b) uncompensated AFMs. The plots depict the eigenvector of $\bm{L}$ (see Eq.~(\ref{eq:LMatrix})) associated to the confined magnon appearing for $K/|J|=0.05, \Delta K_s=0$ in the $d=1$ AFM. The compensated AFM has equal numbers of up and down spins in its ground state, so it can not have a center of inversion symmetry. This lack of inversion symmetry results in two degenerate acoustic confined magnons, one that modulates spin in the left edge (shown) and the other in the right (not shown). In contrast, uncompensated AFM can have an inversion center at the middle spin. This leads to confined magnons that are either symmetric (shown) or antisymmetric (not shown) with respect to inversion at the middle spin}.}
	\label{fig:CompUncomp}
\end{figure}

The sensitivity to different edges is further illustrated by considering the difference between compensated (zero magnetization) and uncompensated AFMs. The former does not have inversion symmetry, while the latter has an inversion center at the middle spin. The edge modes reflect this symmetry, as illustrated in Fig.~\ref{fig:CompUncomp}. 

\section{Conclusions \label{section:conclusions}} 

We presented a theory for spatial confinement of magnons in ferro and antiferromagnetic systems of dimension $d=1,2,3$. For \RdS{semi-infinite systems with a surface or interface with a nonmagnetic material}~we obtained exact analytical expressions for the confined magnon frequency and length scale. These show that extra anisotropy at the surface plays a crucial role in confining magnons, and that confinement only occurs for a certain range of interface anisotropy and in-plane \RdS{(perpendicular to interface)}~propagation wavevector $\bm{q}_{\parallel}$. 

Our theory for \RdS{semi-infinite}~systems was based on exact analytical solution of the classical equations of motion~(\ref{classicalEOM}). \RdS{We achieved this by using trial solutions of two types. The ones that decay exponentially as a function of the distance from the interface allowed us to predict the existence of confined magnons. In contrast, trial solutions that are phase-shifted standing waves allowed us to understand the impact of the interface on the propagating (bulk) magnons. The techniques developed here are generally applicable to other confinement mechanisms such as defect centers and other types of interfaces such as the ones between two different magnetic materials. They also allow the prediction of scattering properties of propagating magnons, e.g. their reflectivity and transmissivity when they scatter off interfaces}. 

Analytic theory was complemented by exact numerical calculations of the spin-scattering function for finite systems, using the quantum Holstein-Primakoff representation~(\ref{eq:HP}). For large $N$ both methods gave identical results for the magnon modes, \RdS{demonstrating that our analytical expressions are indeed exact for semi-infinite systems}.

Apart from expected differences in dispersion, the phenomena of magnon confinement is quite similar for FMs in $d=1,2,3$ and AFMs in $d=2,3$, in that their length scale for confinement Eq.~(\ref{kappa_perp_final}) is identical. 
Surprisingly, the AFMs in $d=1$ have much stronger magnon confinement at the edges. Confined states appear even when $\Delta K_s=0$, a regime that generates no confined states for FMs in $d=1$. The length scale for confinement is substantially different in $d=1$ AFMs (see e.g. Eq.~(\ref{kappaperp1d})). The presence of a confined mode in these systems was not apparent in previous calculations.\cite{Wieser2008,Wieser2009} \RdS{The impact of dimensionality in AFM models is related to the nonexistence of a wavevector $\bm{Q}_{\parallel}$ describing the staggered ground state in $d=1$ (see text below Eq.~(\ref{mixedEOM}))}.

Confined magnons have been detected in ferromagnetic Fe on W(110) using SPEELS.\cite{Prokop2009} Since electron scattering only penetrates a few monolayers it enables measurements of the confined surface magnon for Fe/vacuum (24 monolayer sample in Fig.~3 of Ref.~\onlinecite{Prokop2009}) and for a single Fe monolayer on W(110). The latter yields $\Delta K_s/|J|\approx 0.1$. Our Fig.~\ref{Fe110} shows that our theory is in good agreement with these experiments, leading to the surprising prediction that the surface magnon actually only exists for a finite range of parameters dictated by the regime where Eq.~(\ref{kappa_perp_final}) is positive. In Fig.~\ref{Fe110} this corresponded to a finite wavevector range, $0.30 \pi < q_z a < 1.7\pi$. 

In contrast, we are not aware of experiments detecting confined magnons in antiferromagnets. Table~\ref{TableAFM} gives numerical estimates for the bulk and confined magnon resonances for antiferromagnets MnF$_2$ and FeF$_2$, and for nanoparticles of room temperature multiferroic BiFeO$_3$ (assuming the AFM order is homogeneous instead of cycloidal).\cite{Allen2019, Aupiais2020} The confined magnons are well below the bulk modes in all these cases so they should be observable with spectroscopic probes in the THz frequency range. 

Measurements of $K_s$ are not available for antiferromagnets, so Table~\ref{TableAFM} assumes $K\approx K_s$.
Based on symmetry and microscopic calculations of single ion anisotropy\cite{deSousa2013} we expect $\Delta K_s/J$ to be sizable for these materials. As we show here the confined magnon frequency is quite sensitive to the value of $\Delta K_s=K_s-K$, so its detection enables characterization of this hard-to-measure quantity.

Confined magnons also shed light on the spin order of surfaces, interfaces, and nanostructures in general.\cite{Levy1981} As we show here, when $\Delta K_s < \Delta K_c$ (Eq.~(\ref{DeltaKc})) 
the confined magnon frequency goes to zero and the interface undergoes a ``spin-flip'' phase transition. \RdS{The characteristic length scale for the interface spin texture is expected to be of the same order of magnitude as the magnon confinement length scale $\kappa_{\perp}^{-1}$, see Eq.~(\ref{kappa_perp_final})}.

We also presented a numerical method to compute the confined spectrum for finite systems with arbitrary spin ordering. We presented explicit numerical calculations of the spin-scattering function in simple $d=1$ models to show whether and how confined magnons can be observed with  spectroscopy methods. 

While it is known that localized spin waves (solitons) are present in infinite antiferromagnets when the spin excitations have large amplitude,\cite{Lai1996} the presence of confined modes in the low energy regime relevant for spectroscopy has not been discussed in the literature. 

When $K_s<K$ the interface spins are ``softened'' and in this regime the associated confined magnon lies well below the lowest bulk mode. We predict these confined magnons should be easily detectable by optical probes for both FMs and AFMs as shown in 
Figs.~\ref{fig:FMacousticShoulder}~and~\ref{fig:AFM_KsAsymmetric}. 

For nanostructures with large surface to volume ratio the confined magnon resonance will be a sizable fraction of the bulk $q=0$ peak, even for spatially uniform probes of the spin-scattering function. As shown in Fig.~\ref{fig:AFM_KsAsymmetric} its frequency and spectral weight are quite sensitive to the value of $\Delta K_s$, showing that \RdS{photon scattering}~experiments may successfully probe interface spin anisotropy. 

In conclusion, our work established the crucial role of surface/interface spin anisotropy on driving the emergence of confined magnons. It shows that nanostructures will have several confined magnon resonances in addition to the usual bulk modes often invoked to interpret their magnetic excitations.

\begin{acknowledgments}
S.B. and R.d.S. acknowledge financial support from NSERC (Canada) through its Discovery Program (Grant No. RGPIN-2015-03938 and RGPIN-2020-04328). 
R.F. acknowledges support by the U.S. Department of Energy, 
Office of Basic Energy Sciences, Materials Sciences and Engineering Division.
We thank M. Allen, B.C. Choi, and M. Harder for useful discussions.
\end{acknowledgments}

\bibliography{ConfinedMagnons}

\end{document}